\documentclass[5p,numafflabel]{elsarticle}

\journal{Nature Communications}

\biboptions{numbers,sort&compress,super}

\abstracttitle{Summary}

\usepackage{libertine}
\usepackage{libertinust1math}

\usepackage{amsmath}
\usepackage{bbold}
\usepackage{graphicx}
\usepackage{eurosym}
\usepackage{mathtools}
\usepackage{url}
\usepackage{booktabs}
\usepackage{epstopdf}
\usepackage{xfrac}
\usepackage{tabularx}
\usepackage{bm}
\usepackage{subcaption}
\usepackage{blindtext}
\usepackage{longtable}
\usepackage{multirow}
\usepackage{threeparttable}
\usepackage{pdflscape}
\usepackage[acronym]{glossaries}
\usepackage[export]{adjustbox}
\usepackage[version=4]{mhchem}
\usepackage[colorlinks]{hyperref}
\usepackage[parfill]{parskip}
\usepackage[nameinlink,sort&compress,capitalise,noabbrev]{cleveref}
\usepackage[leftcaption,raggedright]{sidecap}
\usepackage[prependcaption,textsize=footnotesize]{todonotes}

\usepackage{lineno}

\usepackage{siunitx}
\DeclareSIUnit\year{a}
\DeclareSIUnit{\tco}{t_{\ce{CO2}}}
\DeclareSIUnit{\sieuro}{\mbox{\euro}}

\usepackage{lipsum}

\usepackage[resetlabels,labeled]{multibib}

\newcommand{\co}{\ce{CO2}~}

\def\co{CO${}_2${\:}}
\def\coe{CO${}_2$e}

\def\runstandard{decr_13_3H_ws} %

\def\runsensnogreenhy{decr_14_3H_ws}

\def\heatmaplowred{elec_s_4_ec_lc3.0_Co2L0.80_3H_2030_0.13_DF_40export}
\def\heatmapmedred{elec_s_4_ec_lc3.0_Co2L0.30_3H_2030_0.13_DF_40export}
\def\heatmaphighred{elec_s_4_ec_lc3.0_Co2L0.10_3H_2030_0.13_DF_40export}

\graphicspath{
    {results/\runstandard/},
}

\begin{document}

\begin{frontmatter}

	\title{The impact of temporal hydrogen regulation on hydrogen exporters and their domestic energy transition}

	\author[oth,tub,lead]{Leon~Schumm\,\corref{correspondingauthor}}
	\author[ieg,alu]{Hazem~Abdel-Khalek}
	\author[tub]{Tom Brown}
	\author[pik]{Falko Ueckerdt}
	\author[oth]{Michael Sterner}
	\author[uoe]{Maximilian~Parzen}
	\author[unipi]{Davide~Fioriti}

	\cortext[correspondingauthor]{Correspondence: leon1.schumm@oth-regensburg.de}

	\address[oth]{Research Center on Energy Transmission and Storage (FENES), Faculty of Electrical and Information Technology, University of Applied Sciences (OTH) Regensburg, Seybothstr. 2, 93053 Regensburg, Germany}
	\address[tub]{Department of Digital Transformation in Energy Systems, Institute of Energy Technology, Technische Universität Berlin, Fakultät III, Einsteinufer 25 (TA 8), 10587 Berlin, Germany}
	\address[ieg]{Fraunhofer Research Institution for Energy Infrastructures and Geothermal Systems IEG, Gulbener Straße 23, 03046 Cottbus, Germany}
	\address[alu]{Albert-Ludwigs Universität Freiburg, Faculty of Environment and Natural Resources, Tennenbacher Str. 4, 79106 Freiburg im Breisgau, Germany}
	\address[pik]{Potsdam Institute for Climate Impact Research, Telegrafenberg, 14473 Potsdam, Germany}
	\address[uoe]{University of Edinburgh, Institute for Energy Systems, EH9 3DW Edinburgh, United Kingdom}
	\address[unipi]{University of Pisa, Department of Energy Systems, Territory and Construction Engineering, Largo Lucio Lazzarino, 56122 Pisa, Italy}
	\address[lead]{Lead contact}

	\begin{abstract}
		
As global demand for green hydrogen rises, potential hydrogen exporters move into the spotlight.
However, the large-scale installation of on-grid hydrogen electrolysis for export can have profound impacts on domestic energy prices and energy-related emissions.
Our investigation explores the interplay of hydrogen exports, domestic energy transition and temporal hydrogen regulation, employing a sector-coupled energy model in Morocco. 
We find substantial co-benefits of domestic climate change mitigation and hydrogen exports, whereby exports can reduce domestic electricity prices while mitigation reduces hydrogen export prices.
However, increasing hydrogen exports quickly in a system that is still dominated by fossil fuels can substantially raise domestic electricity prices, if green hydrogen production is not regulated.
Surprisingly, temporal matching of hydrogen production lowers domestic electricity cost by up to 31\% while the effect on exporters is minimal. 
This policy instrument can steer the welfare (re-)distribution between hydrogen exporting firms, hydrogen importers, and domestic electricity consumers and hereby increases acceptance among actors.
	\end{abstract}


\end{frontmatter}

%

%

%

%
%
%
%
%
%

%

%

%

%

%

\section*{Introduction}
\label{sec:intro}

The global energy environment is experiencing fundamental upheaval, driven by the need to reduce greenhouse gas emissions and transition to a low-carbon future. In this context, hydrogen has emerged as a viable clean energy carrier capable of addressing the issues of decarbonizing numerous sectors such as industry, transport, heating and power generation. A growing number of countries are investigating green hydrogen production and use as a crucial component of their strategy to cut greenhouse gas emissions and meet ambitious climate targets.
Simultaneously, several countries are positioning themselves as potential exporters of hydrogen and Power-to-X products, discovering an opportunity to leverage their renewable energy resources and technological advances. %
These countries are expected to play a substantial role in the global energy market by supplying clean hydrogen and therefore contributing to global decarbonization efforts. However, pursuing both (on-grid) hydrogen exports and national energy transition raises questions on welfare redistribution, prices and co-benefits that require in-depth analysis, additionally shining a light on the role of temporal hydrogen regulation.

Morocco serves as a blueprint for investigating these dependencies. Morocco, which has vast solar and wind resources \cite{Peters2023, Touili2018, Sterl2022}, has proposed ambitious renewable energy adoption objectives, displaying a commitment to lowering its own greenhouse gas emissions \cite{CAT2021}. At the same time, the country is strategically positioned to deploy its renewable energy potentials to export green hydrogen, opening the door to substantial economic opportunities and local value chains \cite{Ersoy2022}. Furthermore, the country is a net importer of energy \cite{IEA2022}, and green hydrogen could help reduce its dependency on energy imports. The proximity to Europe, which is expected to be a major hydrogen importer, makes Morocco an attractive potential exporter.

These conditions are the motivation of several studies \cite{vanWijk2021, AbouSeada2022, vanderZwaan2021, Schellekens2010, Cavana2021, Touili2022, Timmerberg2019a, Sens2022} examining the potential of hydrogen in Africa and synergies with European demand. A more detailed study of Morocco has been undertaken by Boulakhbar et al.\cite{Boulakhbar2020} examining challenges in integrating Renewable Energies, Khouya et al.\cite{Khouya2020} determining the Levelized Cost of Hydrogen based on concentrated solar power and wind farms, and Touili et al.\cite{Touili2018} investigating the potential of hydrogen from solar energy. Hampp et al.\cite{Hampp2023} investigates various PtX products and their transportation to Europe, and Eichhammer et al. \cite{Eichhammer2019} highlights diverse opportunities and challenges related to exporting hydrogen and Power-to-X products from Morocco. While several studies \cite{Hampp2023, AbouSeada2022, vanWijk2021} have examined the potential of hydrogen as a low-carbon energy carrier and others have explored various countries' climate targets and aspirations \cite{Boulakhbar2020}, a substantial research gap remains regarding the integrated investigation of both perspectives.

Apart from hydrogen exports and domestic climate change mitigation, temporal hydrogen regulation plays a decisive role in prices for domestic consumers and hydrogen exporters. 
Temporal hydrogen regulation defines the rules for green hydrogen production when there is no direct connection between the electrolyser and green electricity production.
Temporal matching in selected European countries is investigated in Zeyen et al.\cite{Zeyen2024}, whereas Ruhnau et al.\cite{Ruhnau2023a} points out benefits of relaxing simultaneity requirements of renewable electricity (RE) supply and hydrogen generation. 
The effect of regulatory options on social welfare and carbon emissions is assessed in Brauer et al.\cite{Brauer2022}. The interplay of additionality criteria and time matching requirements is investigated in Giovanniello et al.\cite{Giovanniello2024}, highlighting how additionality drices the emissions impact of temporal hydrogen regulation.
While Zeyen et al.\cite{Zeyen2024} contextualizes the role of temporal hydrogen regulation with decarbonization scenarios of Germany and Netherlands, none of the mentioned studies fully integrates domestic climate change mitigation and temporal hydrogen regulation scenarios. Furthermore, none of these studies looked into exports in depth.

Numerous studies focus exclusively on either hydrogen exports, domestic climate mitigation or temporal hydrogen regulation, without discussing the complex relationship between these three dimensions. These studies miss interactions between on-grid hydrogen electrolysis and the domestic electricity system, fail to uncover potential synergies and conflicts for both hydrogen exporters as well as domestic electricity consumers with respect to different temporal hydrogen regulation regimes.

The novelty of our study is the development of a sector-coupled energy model for the target region, the inclusion of modelling the temporal hydrogen regulation, and the evaluation of the synergies among hydrogen exports, domestic energy transition and hydrogen regulation.
Additionally, we present a broad scenario vector, sweeping along these three dimensions:
\begin{enumerate}
    \item \textbf{Domestic climate change mitigation}: The domestic climate change mitigation varies between 0--100\% based on Morocco's current emissions of 72 Mt\coe,
    \item \textbf{Hydrogen export}: The hydrogen export volume varies from 1--120 TWh, in accordance with Morocco's hydrogen export ambitions of 114.7~TWh/a and
    \item \textbf{Temporal hydrogen regulation}: The temporal matching (of additional renewable electricity and the electrolyser electricity demand) varies between: no regulation, annual, monthly and hourly matching.
\end{enumerate}
This three-dimensional scenario space results in 264 model runs (excluding sensitivity analysis) to grasp the full extent of the interaction between various parameters.
A 3-hourly resolution is chosen to capture energy system dynamics (e.g. RE generation profiles, energy storage operation) with its diurnal and seasonal variations in energy supply and demand. A similar study \cite{Neumann2022} finds minor underestimation of short-term battery storage and onshore wind and minor overestimation of solar photovoltaics (PV) and hydrogen storage compared to an hourly resolution, overall justifying the reduction of the model size. 
A spatial resolution of 14 nodes represents the geographical heterogeneity of RE resources, demand centers and energy networks. Furthermore, the spatial resolution provides insights into land use conflicts and competing RE resources as well as a spatial differentiation of export ports. Since the 14 nodes align with the Global Administrative Areas level 1 regions, policy advisement can be tailored to specific regions considering domestic needs and constraints.

To investigate interactions between hydrogen exports, domestic climate change mitigation and temporal hydrogen regulation, this research paper presents a fully sector-coupled capacity expansion and dispatch model of Morocco, which includes both gas pipelines and electricity networks. By examining various hydrogen export volumes, climate targets and temporal matchings, we evaluate the potential impact of hydrogen exports on domestic electricity consumers and, vice versa, interactions of domestic climate change mitigation on hydrogen exporters.
Shortcomings of studies on highly renewable energy systems in Africa as low temporal and spatial resolutions as well as the lack of sector-coupled energy models as pointed out in Oyewo et al.\cite{Oyewo2023} are tackled in this study. The analysis of Morocco's energy system contributes to the expanding research that aims to inform policymaking and decision-making regarding sustainable energy transitions. This study provides valuable insights into the pathways unlocking synergies and reducing conflicts between hydrogen exports and national energy transition, enabling Morocco and other potential exporting regions to follow a harmonious and sustainable trajectory while facing similar energy system planning challenges.

Chapter \nameref{sec:intro} presents Morocco's hydrogen strategy and climate targets and outlines potential conflicts and synergies between these two goals. Introducing the results along these three dimensions step-by-step, \nameref{sec:results} evaluates synergies and conflicts based on prices and total cost for domestic electricity consumers and hydrogen exporters, with emphazised consideration of temporal hydrogen regulation. These results are contextualized along policy recommendations in the \nameref{sec:discussion}. \nameref{sec:methods} presents the applied sector-coupled energy model and scenario dimensions required to expose the export-mitigation-regulation nexus.

\subsection*{Hydrogen export strategy and climate targets}

Morocco has implemented various strategies and policies to reach its climate targets and promote hydrogen exports. These include programs on the expansion of RE, the development of hydrogen for domestic demands and exports as well as climate targets \cite{MarHyStrat2021, CAT2021}.

\begin{figure}
    \centering
    \includegraphics[width=\linewidth]{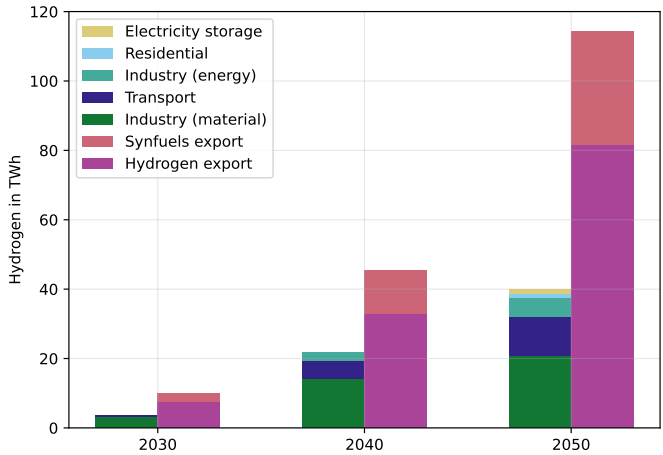}
    \caption{Hydrogen Strategy of Morocco \cite{MarHyStrat2021}. The export volume rises up to 115~TWh/a in 2050, the national demand up to 40 TWh/a in 2050.}
    \label{fig:mar_hydrogen_strategy}
\end{figure}

Figure \ref{fig:mar_hydrogen_strategy} shows the hydrogen strategy of Morocco, including national and export demands.
By 2030, the total hydrogen demand adds up to 13.9 TWh/a and ramps up to 67.9 TWh/a in 2040 and 153.9 TWh/a in 2050, 
clearly listing higher demands for export than for domestic use. The hydrogen generation is backed by 5.2 GW of RE in 2030, 23 GW in 2040, and 57.4 GW in 2050 in the export sector. To cover the national demand of hydrogen, a Renewable Energy deployment of 1.6 GW in 2030, 7.0 GW in 2040 and 10.3 GW in 2050 is planned in Morocco's Hydrogen Strategy\cite{MarHyStrat2021}.

\begin{figure}[h!]
    \centering
    \includegraphics[width=\linewidth]{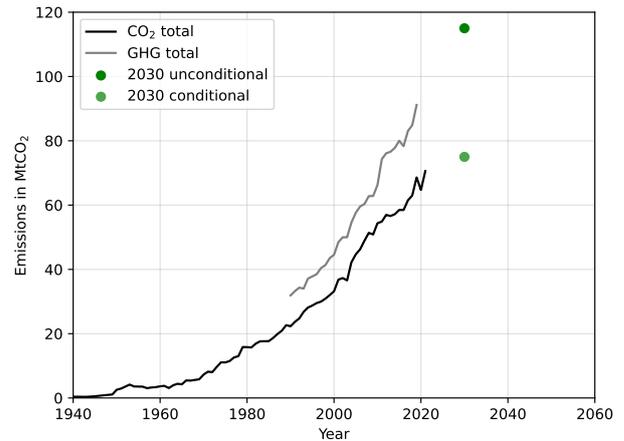}
    \caption{Morocco's historical \co and greenhouse gas emissions (GHG). While emissions are rising, emission targets are in the range between 75 Mt\coe\ (conditional) and 115 Mt\coe\ (unconditional).}
    \label{fig:morocco_em}
\end{figure}

In addition to the hydrogen strategy, Morocco has obliged to climate targets. Figure \ref{fig:morocco_em} displays the historical emissions and targets of Morocco.
According to the self-defined national climate pledges under the Paris Agreement (NDC's), Morocco aims to limit historically rising greenhouse gas emissions to 75 Mt\coe\ (conditional) respectively 115 Mt\coe\ (unconditional) excluding LULUCF by 2030 \cite{CAT2021}. 
As stated by the Climate Action Tracker\cite{CAT2021}, Morocco has not yet submitted a net-zero target.

Both the hydrogen export strategy and climate targets imply a substantial expansion of RE capacities. Furthermore, both strategies require an energy infrastructure development of e.g. electricity grid, hydrogen pipelines, $\mathrm{CO_2}$ network, ports and a considerable scale-up of electrolysers.
Limited resources (e.g. land availability, renewable potentials, workforce, capital) require careful planning to minimize conflicts and maximize possible synergies.

\section*{Results}
\label{sec:results}

\begin{figure*}[h!]
    \centering
    \begin{subfigure}[b]{0.49\linewidth}
        \centering
        \includegraphics[trim={0cm 0cm 0cm 1cm}, clip, width=\linewidth]{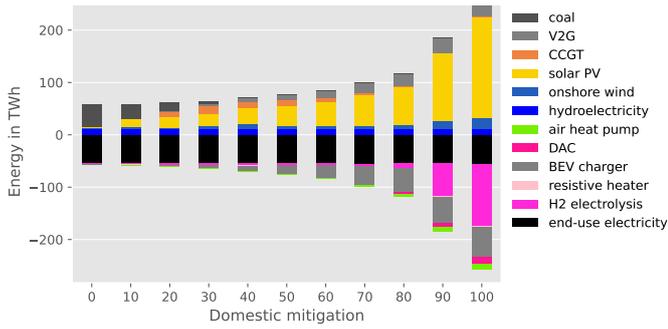}
        \caption{Fixed (1 TWh) export and 0--100\% domestic climate change mitigation}
        \label{fig:balances-ac-0exp-120}
    \end{subfigure}
    \hfill
    \begin{subfigure}[b]{0.49\linewidth}
        \centering
        \includegraphics[trim={0cm 0cm 0cm 1cm}, clip, width=\linewidth]{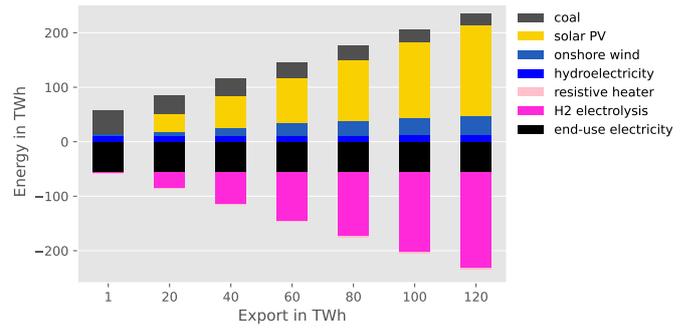}
        \caption{Fixed (0\%) domestic climate change mitigation and 1--120 TWh/a export}
        \label{fig:balances-ac-co2l20-120}
    \end{subfigure}
    \hfill
    \caption{Electricity supply and demand at fixed export levels and increasing domestic climate change mitigation (\ref{fig:balances-ac-0exp-120}) and vice versa (\ref{fig:balances-ac-co2l20-120}). Increasing domestic climate change mitigation first phases out carbon-intensive coal generation in favor of CCGT, at medium to high domestic climate change mitigation the electricity system is fully renewable supported by flexibility through Vehicle-to-Grid (V2G) and sector coupling. Increasing electricity demands include Battery Electric Vehicles (BEV) and hydrogen generation for other sectors.
    At increasing hydrogen exports the additional electricity required for hydrogen electrolysis is covered by onshore wind and solar PV, as imposed by the temporal hydrogen regulation. 
    }
    \label{fig:balances-ac}
\end{figure*}

\subsection*{Domestic climate change mitigation increases the domestic electricity demand}
\label{subsec:increase_limit}
First of all, we investigate the supply and demand of Morocco's electricity system depending on domestic climate change mitigation (dimension 1). Figure \ref{fig:balances-ac-0exp-120} depicts the supply and demand of the electricity system at increasing climate mitigation ambitions from 0\% to 100\%. At low climate ambitions, the electricity demand is mainly covered by coal power, existing (brownfield) capacities of onshore wind, hydro, and combined-cycle gas turbines (CCGT) play only a minor role. 
At increasing domestic climate change mitigation, coal power is phased out in favour of solar PV, furthermore a fuel switch from coal to gas (CCGT) is observable at increasing emission reductions. At medium climate mitigation ambitions, the dipatchable power in the electricity system is provided by CCGT. 
Further increasing the climate ambitions, wind onshore and especially solar PV penetrate and dominate the electricity system complemented by dispatchable power from Vehicle-to-Grid providing an almost fully renewable electricity sector at 100\% emission reduction. 

At 70\% and above, the electricity demand of eletrolysers increases substantially to supply hydrogen allowing a switch from fossil oil products to Fischer-Tropsch fuels in various sectors (s. Fig. \ref{fig:oil-balance}). 
The Fischer-Tropsch demand in the transport sector increases, even though the increasing Battery Electric Vehicle (BEV) diffusion is counterbalancing the demand for oil products (see \nameref{sec:si} section \nameref{subsec:bev_diffusion}). 
Both electrolysers and BEVs drive the electricity demand substantially, up to two times of the domestic end-use electricity demand. 
Additionally, the electricity demand for air heat pumps increases, supporting the defossilisation in the heating sector and supplying heat for Direct Air Capture required for synthetic fuels.

\subsection*{Hydrogen exports require a multiple of the domestic electricity demand}
\label{subsec:increase_h2}

Here, we analyse the hydrogen export ramp up (dimension 2). The electricity supply at increasing hydrogen exports is displayed in Figure \ref{fig:balances-ac-co2l20-120}. The deployment of hydrogen exports requires a scale up of onshore wind and solar PV accordingly, as defined by the green hydrogen constraint outlined in \nameref{subsec:green_hydrogen_constraint}.
Without the temporal hydrogen regulation, the additional electricity demand of hydrogen exports is covered by coal power plants and CCGT (by increasing the capacity factor of existing brownfield coal and CCGT capacities) and an expansion of open-cycle gas turbines (OCGT) installation and supply as pointed out in Figure \ref{fig:barplotscons}.
In contrast to the exponential increase of electricity demand observable at increasing climate mitigation ambitions displayed in Figure \ref{fig:balances-ac-0exp-120}, the electricity demand increases linearly with the hydrogen export ambitions.

\begin{figure*}[h!]
    \centering
    \begin{subfigure}[b]{0.49\linewidth}
        \centering
        \includegraphics[width=\linewidth]{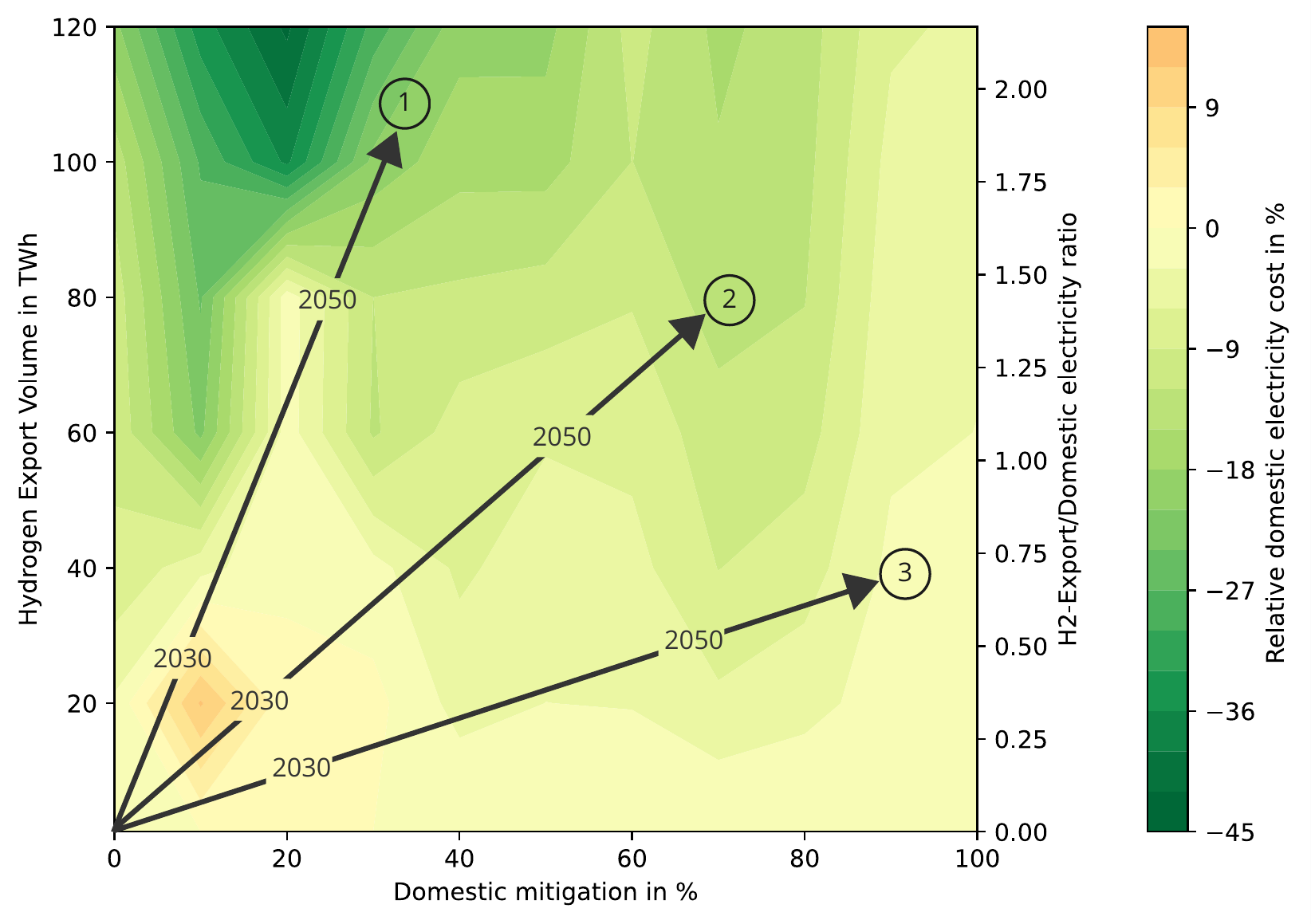}
        \caption{Relative cost of electricity for domestic customers (normalized to 1 TWh/a hydrogen export)}
        \label{fig:expense_ac_120}
    \end{subfigure}
    \hfill
    \begin{subfigure}[b]{0.49\linewidth}
        \centering
        \includegraphics[width=\linewidth]{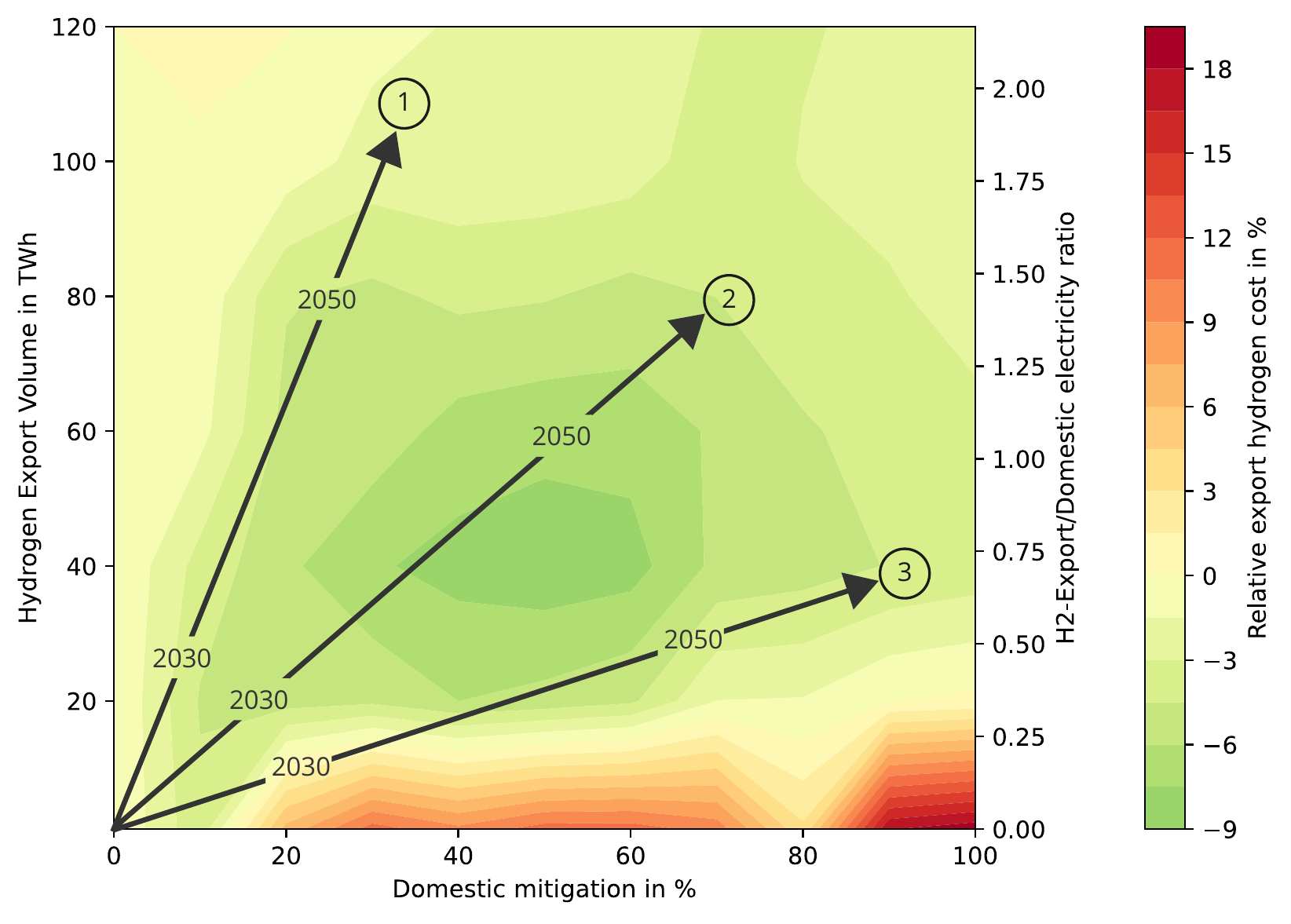}
        \caption{Relative cost of hydrogen for exporters (normalized to 0\% domestic climate change mitigation)}
        \label{fig:expense_h2_120}
    \end{subfigure}
    \hfill
    \caption{  
    Cost for domestic electricity consumers (\ref{fig:expense_ac_120}) and hydrogen exporters (\ref{fig:expense_h2_120}),
    normalized to costs at 1 TWh/a hydrogen export (\ref{fig:expense_ac_120}) and
    to 0\% \co reduction (\ref{fig:expense_h2_120})
    at each domestic climate change mitigation level. Domestic electricity consumers profit from increasing hydrogen exports, especially at low domestic climate change mitigation and high exports. Hydrogen exporters profit from domestic climate change mitigation at medium mitigation efforts. Both (\ref{fig:expense_ac_120}) and (\ref{fig:expense_h2_120}) include possible pathways of i) quick exports and slow climate change mitigation, ii) balanced exports and mitigation and iii) slow exports and quick climate change mitigation. Years are illustrative.}
    \label{fig:expenses_default_120}
\end{figure*}

\subsection*{Domestic climate change mitigation and hydrogen export show co-benefits}
\label{subsec:benefits}

In an integrated analysis of domestic climate change mitigation (dimension 1) and hydrogen exports (dimension 2) is study outlines that both domestic climate change mitigation and hydrogen exports profit from each other and show co-benefits. In this Section, we i) show the effects of hydrogen exports on domestic electricity prices, ii) the role of domestic climate change mitigation on hydrogen export cost, and lastly iii) common co-benefits.

First, Figure \ref{fig:expense_ac_120} shows the cost for domestic electricity consumers subject to mitigation and hydrogen export volumes. The costs are normalized to costs at 1 TWh/a Hydrogen export at each domestic climate change mitigation level, displaying relative changes for domestic electricity consumers induced by hydrogen exports at a certain domestic climate change mitigation.
Hydrogen exports decrease the relative domestic electricity cost by up to 45\%, especially at mitigation below 40\% and exports above 50 TWh. In these ranges, the fossil dominated domestic electricity system profits from excess green electricity originating from additional renewable energy capacities required for hydrogen export. The more the domestic electricity system is decarbonized (domestic climate change mitigation above 50--60\%), the weaker is the decrease of domestic electricity prices with hydrogen exports.
The only increase of cost induced by hydrogen exports is observable at a low climate change mitigation of 10\% and exports of 20 TWh. 

Second, Figure \ref{fig:expense_h2_120} shows the cost of hydrogen exports subject to mitigation and hydrogen export volumes. The costs are normalized to costs at 0\% climate change mitigation at each hydrogen export volume, displaying relative changes for hydrogen exports from domestic climate change mitigation at a certain hydrogen export volume.
At hydrogen export volumes of 25--75 TWh, an increase of domestic climate change mitigation up to 40--60\% decreases the hydrogen export prices up to -9\%. 
The only price increase observable in the range of our scenarios is observable at low export volumes below 20 TWh. Here, advances in domestic climate change mitigation increase the hydrogen export cost by up to 19\% compared to no domestic climate change mitigation.
This cost increase results from lower electrolysis capacity factors compared to the 0\% climate change mitigation scenario, where we observe high electrolysis capacity factors of above 80\% (s. Fig. \ref{fig:cf-ely}).
The triangle in the top left area shows that cost for hydrogen exporters at certain export levels are independent of domestic climate change mitigation, this is driven by temporal hydrogen regulation further investigated in section \nameref{subsec:benefits_rule}. Here, hydrogen export infrastructure decouples from the domestic electricity system, the cost is increasingly independent of domestic climate change mitigation. The hydrogen exports are 2 times higher than the domestic electricity demand (s. Fig. \ref{fig:expense_h2_120}), dwarfing the importance of the domestic electricity system.

Third, we derive co-benefits for domestic climate change mitigation and hydrogen exports. 
Within the area of 40-60\% climate change mitigation and 50-100 TWh/a hydrogen exports, we see i) domestic electricity consumers profit from hydrogen exports compared to very low (1 TWh) exports and ii) hydrogen exporters decrease their cost compared to no (0\%) domestic climate change mitigation. Hence, both domestic climate change mitigation and hydrogen exports show clear co-benefits at medium mitigation (40--60\%) and moderate to high hydrogen exports (25--120 TWh).

Apart from the cost of electricity for domestic customers and cost of hydrogen for exporters, Figure \ref{fig:expenses_default_120} presents possible pathways of climate change mitigation and export. We have derived three illustrative mitigation-export pathways presenting various speeds of transformation:
\begin{enumerate}
    \item \textbf{Quick exports and slow climate change mitigation},
    \item \textbf{Balanced exports and climate change mitigation} and
    \item \textbf{Slow exports and quick climate change mitigation}.
\end{enumerate}

In the long run, domestic electricity consumers profit in all pathways from hydrogen exports. A slight and short increase of domestic electricity prices in the early transformation phase (2030) is compensated by later (2050) strong (pathway 1) and medium (pathway 2) benefits for domestic electricity customers, profiting from hydrogen exports.
In pathway 3, the short-term price increases are only marginal, but the long-term benefit from decreased is not as strong as in the scenarios with quicker export scale-ups. Figure \ref{fig:expense_h2_120} shows that hydrogen exporters have clear benefits in pathway 2, the cost of hydrogen for exports decrease compared to no domestic climate change mitigation. By ramping up hydrogen exports in an early stage of domestic climate change mitigation, high relative cost of hydrogen can be avoided.

\begin{figure}[h!]
    \centering
    \includegraphics[trim={0cm 0cm 0cm 0.65cm}, clip, width=\linewidth]{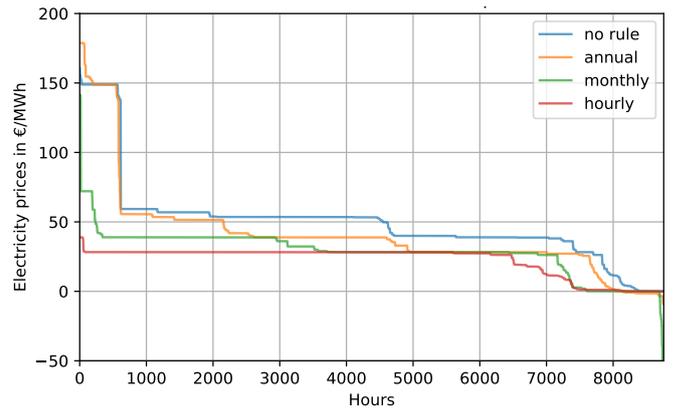}
    \caption{Price duration curve at 120 TWh/a export and 0\% domestic climate change mitigation. Stricter temporal hydrogen regulation pushes the price duration curve towards the left, since additional renewable electricity capacities phase out fossil generation with higher marginal costs than renewables. Negative prices below -50 €/MWh are cut off.}
    \label{fig:pdc-120-0}
\end{figure}

\begin{figure*}[h!]
    \centering
    \begin{subfigure}[b]{0.49\linewidth}
        \centering
        \includegraphics[trim={0cm 0cm 0cm 0.65cm}, clip, width=\linewidth]{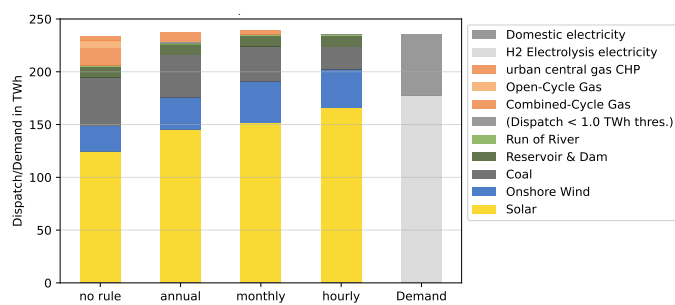}
        \caption{Electricity dispatch and demand}
        \label{fig:dispatch_rule}
    \end{subfigure}
    \hfill
    \begin{subfigure}[b]{0.49\linewidth}
        \centering
        \includegraphics[trim={0cm 0cm 0cm 0.65cm}, clip, width=\linewidth]{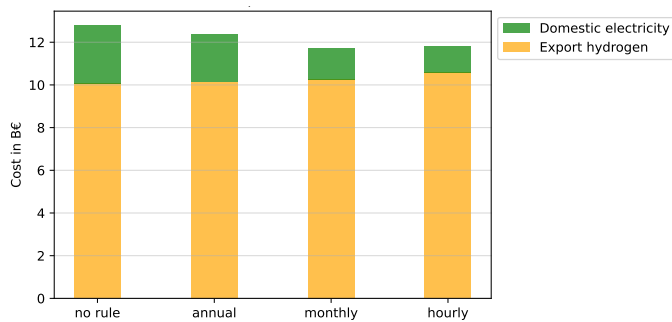}
        \caption{Cost for domestic electricity consumers and hydrogen exporters}
        \label{fig:expense_h2ac}
    \end{subfigure}
    \hfill
    \caption{Electricity dispatch and demand (\ref{fig:dispatch_rule}) and cost for consumers (\ref{fig:expense_h2ac}) for various (hydrogen) temporal matching regimes in the 120 TWh/a export and 0\% climate change mitigation scenario. Stricter temporal matching decreases carbon-intensive electricity generation (coal \& gas) for hydrogen generation and even domestic electricity consumers (s. Fig. \ref{fig:dispatch_rule}). Cost for export hydrogen generation increase to fulfill the temporal matching constraint, whereas domestic electricity consumers profit from stricter hydrogen regulation.}
    \label{fig:expenses_rule}
\end{figure*}

\subsection*{Hydrogen regulation reduces domestic electricity prices and emissions}
\label{subsec:benefits_rule}

Hydrogen regulation has strong effects on electricity system dispatch (and emissions), increasing the prices and cost for hydrogen exporters whilst decreasing for domestic electricity consumers. This section shows the mechanisms of temporal matching in the 120 TWh/a export and 0\% domestic climate change mitigation scenario.

As the temporal matching becomes stricter, additional renewable capacities and complementing hydrogen storage are required to meet the electricity demand of electrolysis on an annual, monthly and hourly basis, as shown in Figure \ref{fig:tsc-120-0}. These renewable capacities i.) push out high marginal cost fossil generation (\textit{price spillover}) and ii.) provide excess electricity to the domestic electricity system (\textit{energy spillover}). Both spillover effects are linked, the energy spillover causes a merit order effect affecting the domestic electricity prices.

The \textit{price spillover} is observable in Figure \ref{fig:pdc-120-0} depicting the price duration curve at 120 TWh/a export and 0\% domestic climate change mitigation. The price duration curve describes the sorted price of electricity (spatial mean) at every hour of the year.
Stricter hydrogen regulation pushes the price duration curve towards the left, since additional renewable electricity capacities phase out fossil generation with higher marginal costs than renewable electricity.

The \textit{energy spillover} is shown in Figure \ref{fig:dispatch_rule}. In the \textit{no rule} scenario, which does not constrain the electricity input of hydrogen electrolysis, coal and OCGT are running close to their (brownfield) capacity limits to supply electricity. If annual matching is applied, the renewable electricity supply to the electricity system equals the annual demand for electrolysis, increasing the supply of solar PV by 20 TWh/a while decreasing the supply of fossil electricity generation. 
Monthly matching increases this trend, while hourly matching completely phases out OCGT and CCGT. Figure \ref{fig:dispatch_rule} highlights the \textit{energy spillover} effect: going beyond the electricity demand of hydrogen electrolysis, renewable capacities installations necessary to meet the temporal matching decarbonize the domestic electricity demand and decrease the grid emissions. This effect is steered by the strictness of temporal matching, hence stricter temporal matching rules enable a further decrease in emissions.
The slight increase in total dispatch for annual and monthly matching is balanced by resistive heaters in the heating sector.
In addition to the synergies between hydrogen export and domestic climate change mitigation, the \textit{energy spillover} outlines a clear synergy between domestic climate change mitigation and temporal hydrogen regulation.

The \textit{price spillover} effects results in lower electricity prices for domestic consumers. Taking the electricity demand into account, the decrease of prices results in a decrease of electricity cost shown in 
Figure \ref{fig:expense_h2ac}. Stricter temporal hydrogen regulation decreases the domestic electricity cost from 2.7 B€ to 1.2 B€. In contrast, the hydrogen cost for exporters increases from 10.1 B€ to 10.6 B€, carrying the cost of the temporal matching constraint and hence financing additional solar PV and hydrogen storage required to meet the electrolysis electricity demand on a certain temporal basis.

The combined cost of domestic electricity and export hydrogen decreases with stricter temporal hydrogen regulation, whereas the total system cost increase as displayed in Figure \ref{fig:tsc-120-0}. This seeming inconsistency is based on the fact, that electricity consumers (domestic and hydrogen electrolyser input) pay less contribution margin to the cost of capital of coal power plants, since they get phased out with stricter temporal hydrogen regulation. On the other hand, the cost of capital of existing coal power plants is not part of the optimization-based total system cost, since they are a brownfield capacity and not extended. In short, the reduction of utilisation of the brownfield coal power assets decreases the cost for electricity consumers, whereas the additional renewable capacities required to meet the temporal hydrogen regulation drive the total system cost as shown in Figure \ref{fig:tsc-120-0}.

To sum up, temporal hydrogen regulation is a policy instrument acting in a twofold way. It steers the redistribution between hydrogen exports and domestic electricity consumers (based on the \textit{price spillover} effect), decreasing the cost for domestic electricity consumers and increasing the cost for hydrogen exporters. The \textit{energy spillover} effect not only decarbonizes hydrogen exports but decreases grid emissions for  domestic electricity demand as well. Both effects scale with stricter temporal matching.

\subsection*{The effects of temporal hydrogen regulation are strongest in high export and low climate change mitigation systems}
\label{subsec:rule_all}

\begin{figure}[h!]
    \centering
    \includegraphics[trim={0cm 0cm 0cm 0.65cm}, clip, width=\linewidth]{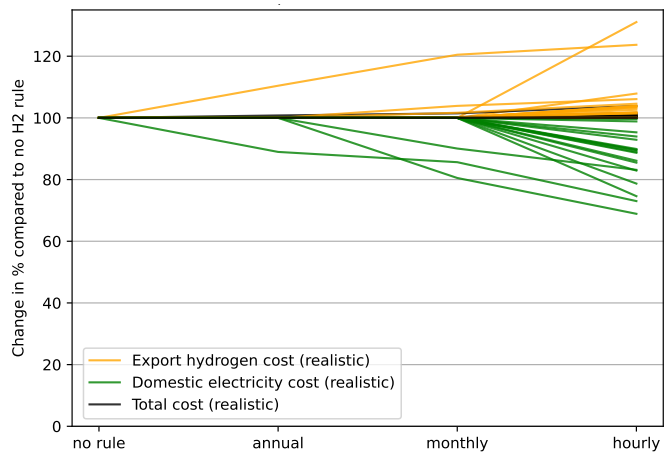}
    \caption{Relative change of electricity and hydrogen cost and total system cost depending on the temporal hydrogen regulation. Domestic electricity consumers profit across all export and mitigation scenarios but most at high export and low climate change mitigation. Hydrogen exporters experience higher cost with stricter temporal hydrogen regulation. The temporal hydrogen regulation regulates the welfare distribution between both groups.}
    \label{fig:expenses_real_120}
\end{figure}

Based on the effects of temporal hydrogen regulation in the scenario of 120 TWh/a export and 0\% climate change mitigation presented in the previous section, this section explores the temporal hydrogen regulation across the pathway ii.) \textit{Balanced exports and mitigation} by combining all three dimensions of domestic climate change mitigation, hydrogen export and temporal hydrogen regulation. Figure \ref{fig:expenses_real_120} shows the relative change of electricity and hydrogen cost of the \textit{balanced exports and mitigation} scenarios in dependence of temporal matching. 
Across all displayed scenarios, temporal hydrogen regulation hedges domestic electricity consumers against rising prices. Depending on the domestic climate change mitigation and hydrogen export, hourly matching decreases the cost of electricity for domestic consumers by up to 31\%.

The scenarios of the \textit{balanced exports and mitigation} pathway in Figure \ref{fig:expenses_real_120} show that in only three mitigation-export combinations the annual or monthly matching decreases the cost for domestic electricity, whereas all other combinations of the \textit{Balanced exports and mitigation} pathway are only sensitive to the strictest -- hourly matching -- hydrogen regulation. 

In only two scenarios, hourly matching regulation increases the hydrogen cost above 7\% compared to no temporal hydrogen regulation. This is the case for 0\% climate change mitigation and 1 and 20 TWh/a export scenarios.

The effects of temporal hydrogen regulation on domestic electricity prices are most dominant in high export and low climate change mitigation scenarios. 
In these cases, the introduction of annual matching has already a striking effect, because the \textit{price spillover} observed in Figure \ref{fig:pdc-120-0} decreases (domestic) electricity prices. The \textit{price spillover} unfolds in fossil-dominated (hence low domestic climate change mitigation) systems, combined with high hydrogen exports providing a strong \textit{energy spillover} effect.
See also Figure \ref{fig:expenses_all_200} for  scenarios following the \textit{quick exports and slow climate change mitigation} pathway in which the effect of temporal hydrogen regulation is even stronger.

In contrast, in the \textit{slow exports and quick climate change mitigation} scenarios only the hourly matching has observable effects, since the electricity system is already  dominated by renewable electricity and only hourly matching phases out the remaining fossil generation (s. Fig. \ref{fig:expenses_all_200}). The differentiation among the domestic climate change mitigation leads to these staged effects of temporal hydrogen regulation on domestic electricity cost.

\section*{Discussion}
\label{sec:discussion}

\subsection*{Temporal hydrogen regulation and the speed of transformation}
\label{subsec:timepath}
The transition towards climate neutrality of Morocco's energy system and the ramping up of exports up to 120 TWh/a are a matter of decade(s) and undergo a certain pathway. Based on the co-benefits identified in \nameref{subsec:benefits}, steering the transition by minimizing conflicts and amplifying co-benefits potentially reduces both the price of electricity and price of hydrogen. 

The broad scenario design of this study allows conclusions and recommendations beyond the specific Morocco case.
Various potential hydrogen exporting countries with different prerequisite can be found in the integrated analysis of benefits and burdens for domestic electricity consumers and hydrogen exporters among the dimensions of domestic climate change mitigation, hydrogen export and hydrogen regulation.

Based on the main findings of co-benefits and temporal hydrogen regulation, we present three possible pathways of reaching the aims of hydrogen exports and domestic climate change mitigation complemented with tailored temporal hydrogen regulation recommendations:
\begin{enumerate}
    \item \textbf{Quick exports and slow climate change mitigation}: Reaching high export levels before adequately mitigating domestic emissions leads to higher total emissions, if no strict temporal hydrogen regulation is in place. Instead, hourly matching reduces domestic electricity cost effectively and redistributes welfare from exporters to domestic consumers. This regulation plays a decisive role in this pathway, even annual or monthly matching triggers the \textit{price spillover} and \textit{energy spillover} effects.
    \item \textbf{Balanced exports and climate change mitigation}: A medium (or balanced) transition of both exports and mitigation offers benefits for both domestic electricity consumers and hydrogen exporters. Domestic electricity consumers experience a short phase of higher prices (10\% climate change mitigation and 20 TWh/a export), but are then rewarded towards higher export levels given strict hourly temporal hydrogen regulation.
    \item \textbf{Slow exports and quick climate change mitigation}: If the transition speed of domestic climate change mitigation is quicker than the scale-up of hydrogen exports, this imposes increasing cost for hydrogen exporters, while the effect of exports on domestic electricity cost is marginal. The temporal hydrogen regulation plays only a minor role in this pathway, since the electricity system is already highly or fully renewable electricity based and neither \textit{price spillover} nor \textit{energy spillover} are effective.
\end{enumerate}

The three illustrative pathways offer certain room for manoeuvre in policy-making for various countries and their export ambitions as well as speed of domestic climate change mitigation. However, these pathways do not reflect additional constraints as limited renewable energy ramp-up rates, the availability of workforce, additional prerequisites such as an effective carbon pricing system for carbon-based Power-to-X products and bureaucratic hurdles among others. A holistic approach and well-balanced policies is required to steer the energy transition and export ambitions.

\subsection*{Balancing interests of exporters and domestic demand}
\label{subsec:balancedinterests}
The implications of hydrogen export on domestic electricity consumers are reflected in various dimensions. 

First, the price of electricity for domestic businesses and households need to be considered to enable a fair discussion and balance of burdens and opportunities. This study shows that the price of electricity highly depends on hydrogen export volumes and temporal hydrogen regulation. The export of hydrogen in combination with strict temporal hydrogen regulation provides the opportunity of decreasing domestic electricity prices, providing opportunities for domestic businesses and households.

Second, the substantial deployment of solar PV and onshore wind comes along with a substantial land use. Several studies \cite{Terrapon-Pfaff2019, Hanger2016} highlight the requirement of local acceptance of renewable energies in Morocco. From a German/European perspective, hydrogen imports are not necessarily the most economic option \cite{Merten2023} but also provide a way out of having to deal with the domestic acceptance of renewable energies.
Shifting such issues of acceptance to Morocco by deploying renewable energy and scaling up their hydrogen exports raises moral questions and should be dealt with carefully. 

Third, this study shows the benefits of integrating solar PV and onshore wind into the main grid for both domestic electricity consumers and hydrogen exporters. By integrating the best sites of solar PV and onshore wind, domestic electricity consumers can access the cost-cutting potentials and benefit from lower electricity prices as shown in our results. On the other hand, off-grid hydrogen generation potentially reserves the promising electricity potentials for hydrogen export only. Nonetheless, Tries et al.\cite{Tries2023b} show the benefits of hydrogen islanding, offering cost savings for inverters and benefits for power quality. Domestic consumers may profit from these economic opportunities for hydrogen exporters indirectly through corresponding policies.

Hydrogen export ambitions (fostered by e.g. the European Union) do not cause disadvantages for the domestic population \textit{per se}, a proactive transition taking into account the interplay of domestic electricity prices unleashes economic opportunities for both exporters and domestic population.

\subsection*{Limitations}
\label{subsec:limitations}

The sector-coupled energy model as well as the results underly certain limitations. In this study, we leave the option open that the hydrogen for export is further synthesized to hydrogen derivatives as ammonia, Fischer-Tropsch products or sponge iron etc. We do not model the system operation of a further synthesis of export products explicitly, unlike we do for the domestic demands for hydrogen derivatives. Further synthesis of export products and linked transportation modes are investigated in Hampp et al.\cite{Hampp2023}, Galimova et al.\cite{Galimova2023} and Verpoort et al.\cite{Verpoort2023} but not in the scope of this study.

A further synthesis will likely have effects on the system operation, since the synthesis in combination with cheaper storage (e.g. oil storage is cheaper than hydrogen storage \cite{DEA2019TechnologyData}) provides an additional flexibility to a certain extent but is limited by the operation flexibility of Fischer-Tropsch or Haber-Bosch processes.

The constant hydrogen export demand assumed in this study represents pipeline operation or a further hydrogen synthesis.
If the export hydrogen is not further synthesized but exported as liquified hydrogen via shipping, there are significant impacts on the energy system to expect. The demand pattern of ship export (landing, loading, travel time) triggers spikes in hydrogen demand, which would need to be buffered by large-scale hydrogen storage or by a corresponding operation of hydrogen electrolysis.

Additionally, the transition of certain sectors is not subject to the optimisation but exogenously defined. Such nodal shifts (e.g. in transport: share of electric vehicles) provide system benefits (e.g. Vehicle-to-Grid) for which an integrated optimisation favours e.g. higher shares of electric vehicles.
Recent approaches as Zeyen et al.\cite{Zeyen2023} improve such caveats by incorporating endogenous learning, but research gaps on the endogenous demand of the transport sector remain.

National climate ambitions and especially the scale-up of hydrogen export require substantial amounts of raw materials (e.g. copper, cement) linked with resource limitations as well as upstream greenhouse gas emissions. As pointed out in Wang et al.\cite{Wang2023} in a global analysis, most raw material limitations required for electricity generation do not exceed geological reserves. However, material demands of e.g. electrolysers are not within the scope of Wang et al.\cite{Wang2023}.

The cost of water through desalination and transport has a single cost for the whole country. However, the transportation costs highly depend on the terrain as pointed out by Caldera et al.\cite{Caldera2016}, hence the water costs are higher in more remote areas which is not taken into account in this study. In contrast, the effect of a higher spatial resolution of water costs is assumed to be minor, since the water desalination and transport costs do not contribute substantially to the cost of hydrogen as emphasized in Hampp et al.\cite{Hampp2023}.

\label{sec:conclusion}

\subsection*{Conclusion}

Our study shows the importance of integrated scenario analysis combining hydrogen exports, domestic climate change mitigation and temporal hydrogen regulation.

Both hydrogen exports and domestic climate change mitigation benefit from each other mainly at emissions reduction of 40--60\% compared to the baseline scenario and moderate to high hydrogen exports (25--120 TWh/a).
However, we show that there are also risks with respect to rising domestic electricity prices
and find that hydrogen regulation via temporal matching is a decisive instrument
to protect domestic consumers. Hourly matching decreases the cost for domestic electricity consumers (up to -31\%) while the effect on hydrogen exports is minimal. Temporal hydrogen regulation can effectively hedge domestic electricity consumers against rising prices across mitigation and export scenarios.

This study contributes to the investigation of implications and benefits of hydrogen exports for domestic electricity consumers. Even though temporal hydrogen regulation hedges domestic electricity consumers against rising prices, further implications of hydrogen exporters on the domestic population (land use, competing RE resources, environmental concerns of desalination) are not within the scope of this study.

In summary, our study underscores that hydrogen export ambitions, as encouraged by entities like the European Union, is in favor of domestic electricity consumers at certain export quantities, if the appropriate hydrogen regulation is in place. A proactive and comprehensive transition strategy, considering the interplay between hydrogen exports and domestic climate change mitigation, is key to unlocking economic opportunities for both hydrogen exporters and the domestic population.

Apart from reducing greenhouse gas emissions, temporal hydrogen regulation is a decisive policy instrument in steering the welfare (re-)distribution between i) hydrogen exporting firms and hydrogen importing countries and ii) hydrogen exporting firms and domestic electricity consumers. 
A fair balance can increase acceptance across actors, is crucial for a sustainable energy transition in Morocco and offers valuable insights for further countries facing similar energy challenges globally.

\section*{Methods}
\label{sec:methods}

\subsection*{Sector-coupled energy model}
\label{subsec:moroccan_model}
The sector-coupled energy model of Morocco is based on the global electricity model  PyPSA-Earth \cite{Parzen2023} and the sector-coupling extension \cite{Abdel-Khalek2024} using linear optimisation and overnight scenarios. This capacity extension model optimises the generation, transmission and storage capacities of Morocco's energy system by minimising the the total annualised system costs constrained by e.g. climate targets or green hydrogen policies.

\subsubsection*{Energy demand}
To obtain the annual energy demand, we use the workflow presented by Abdel-Khalek et al.\cite{Abdel-Khalek2024}. In a first step, the annual energy demand is obtained from the United Nations Statistics Database\cite{unstats2023}. The latest available data is from 2020, since these energy balances are likely to show irregularities due to COVID-19 implications, the base year in this study is 2019. Based on the annual energy demand from 2019, the energy demand of 2035 is projected according to efficiency gains and activity growth rates similar to M{\"u}ller et al.\cite{Muller2023}.
In a second step, the annual and national energy demand is distributed according to the production sites of Morocco.
In case of subsectors where no spatially resolved production sites are available (e.g. paper industry), the demand is distributed according to GDP.
Third, the annual but locationally resolved energy demand is converted to hourly demands. In this study, a constant hourly demand of the industry and agriculture sector is assumed whereas electricity, heating, transport are temporally resolved based on time series derived from Brown et al.\cite{Brown2018a}. The resulting temporal and spatial energy demands of the sectors:
\begin{itemize}
    \item Electricity,
    \item Heating,
    \item Industry,
    \item Transport (incl. aviation and shipping) and
    \item Agriculture
\end{itemize}
are integrated in the sector coupled energy model. The inland energy demand per sector remains constant throughout the scenarios presented in the \nameref{sec:intro}, regardless of endogenous hydrogen export volumes or national carbon emission reductions. However, to account for increasing shares of battery electric vehicles (BEV) in the Moroccan energy system, the share of BEV's is linked to the emission reduction ambitions (s. Fig. \ref{fig:bev_diffusion}) from 2\% (today's levels) up to a share of 88\% at 100\% domestic climate change mitigation in accordance with Rim et al.\cite{Rim2021}.

\subsubsection*{Renewable Energy Sources}
The sector-coupled energy model follows the data pipeline of \cite{Parzen2023} by incorporating solar PV, onshore wind and hydro power using the open source package Atlite \cite{Hofmann2021}.
Atlite obtains technology-specific time series based on weather data (ERA5 reanalysis data \cite{Hersbach2020}, SARAH-2 satellite data \cite{Pfeifroth2017}). 
In addition to the time series, Atlite calculates land availabilities and linked potentials using the Copernicus Global Land Service \cite{Buchhorn2020}.

\begin{figure*}[t]
    \centering
    \includegraphics[width=0.7\linewidth]{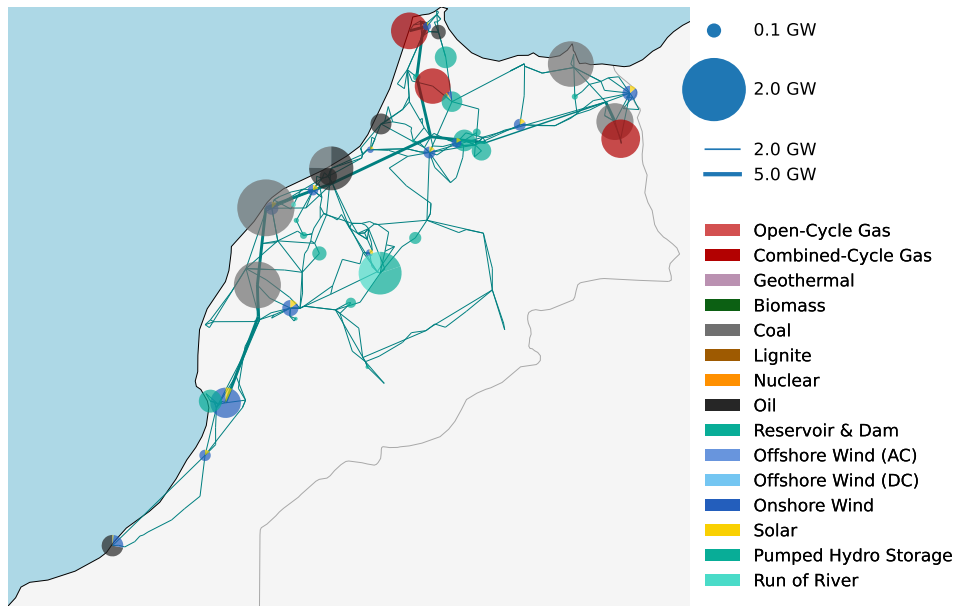}
    \caption{Current capacities of electricity generation and distribution, obtained from \cite{Parzen2022} and visualization based on \cite{Horsch2018}. Morocco's electricity generation portfolio is currently dominated by fossil generation (coal and gas), includes some hydropower plants and increasing but still minor capacities of onshore wind and solar PV. Boundaries depicted are based on the Global Administrative Areas and are intended for illustrative purposes only, not implying territorial claims.
    }
    \label{fig:MAR_brownfield}
\end{figure*}

\subsubsection*{Conventional electricity generation}
The conventional electricity generators are obtained from the \textit{powerplantmatching} tool \cite{Powerplantmatching2019}. \textit{Powerplantmatching} uses various input sources as OpenStreetMap2022, Global Energy Monitor, IRENA \cite{IRENA2022, OpenStreetMap2022, GlobalEnergyMonitor} to download, filter and merge the datasets. 
In this study, \textit{powerplantmatching} is applied to Morocco delivering the conventional power plants.

\subsubsection*{Electricity networks and gas pipelines}
The electricity network is based on the PyPSA-Earth workflow using OpenStreetMap2022 data \cite{OpenStreetMap2022} as presented in Parzen et al.\cite{Parzen2023}. First, the data is downloaded, filtered and cleaned. Second, a meshed network dataset including transformers, substations, converters as well as HVAC and HVDC components is build\cite{Parzen2023}. 
The brownfield gas pipeline infrastructure of Morocco is not considered in this study. The Maghreb-Europe pipeline passing through Morocco is the only pipeline in operation but currently subject to political disputes \cite{Rachidi2022}. Further proposed projects are excluded due to  miniscule capacity compared to Morocco's energy system (Tendrara LNG Terminal, 100 million cubic meters per year or uncertain commissioning dates (Nigeria-Morocco Pipeline, start year 2046 \cite{GEM2023b}).

\subsection*{Brownfield model and capacity expansion}
\label{brownfield_model}
Following the workflow of \cite{Parzen2023}, Figure \ref{fig:MAR_brownfield} shows the current capacities of electricity generation and distribution of Morocco. Based on this brownfield electricity system, the sector-coupled energy model allows the capacity expansion of storage, (electricity networks), hydrogen pipelines, renewable and conventional generators.

\subsection*{Technology and cost assumptions}
\label{subsec:tech_assump}
All technology and cost assumptions are based on the year 2030, taken from \href{https://github.com/pypsa/technologydata}{github.com/pypsa/technologydata} version 0.4.0. A discount rate of 13\% is applied.

\subsection*{Water supply}
\label{subsec:water_supply}
Hydrogen production through electrolysis requires fresh water, alternatives as the direct use of high saline water sources are still in the experimental stages \cite{Tong2020}. The depletion of freshwater resources and it's competition with other water uses is a concern, especially in regions such as Morocco which is ranked 27th among the world's most water-stressed countries \cite{Maddocks2015}. 
Sea Water Reverse Osmosis (SWRO) emerges as a feasible solution to address this issue. In this model, the additional water cost of 0.80 €/\si{\cubic\metre} through desalination and transport is considered in line with the base scenario for Morocco in 2030 in Caldera et al.\cite{Caldera2020}. These findings are comparable to Kettani et al.\cite{Kettani2020} and Caldera et al.\cite{Caldera2016} stating the water costs of 1\$/$m^3$ resp. 0.60 - 1.50 €/\si{\cubic\metre} for Morocco in 2030. Given a water demand of 9~$m^3/kg_{hydrogen}$
\cite{Hampp2023}, the additional costs for electrolysers result in 0.216~$MWh/kg_{hydrogen}$ (LHV).

However, the challenge of brine disposal presents environmental concerns due to treatment chemicals and high salinity as discussed by \cite{Thomann2022, Dresp2019, Tonelli2023} that require a careful site selection of desalination plants. A minimum distance of four kilometers from marine protected areas is recommended in \cite{Thomann2022}.

\subsection*{Green hydrogen policy}
\label{subsec:green_hydrogen_constraint}

A key constraint on the model is that the hydrogen exported from Morocco requires to meet sustainability criteria ("green" hydrogen). A variety of studies looks into various dimensions (temporal, geographical, electricity origin) of green hydrogen and their trade-offs \cite{Brauer2022, Ruhnau2022, Zeyen2024}.
Envisioning hydrogen offtakers from the European Union, the green hydrogen constraint applied in this study aligns with the \emph{Delegated regulation on Union methodology for RFNBOs} of the European Commission \cite{Commission2023} defining green hydrogen  based on the following criteria:

\begin{enumerate}
    \item Additionality: The electricity demand of electrolysers must be provided by additional RE power generation (less than 3 years before the installation of electrolysers, from 1.1.2028 onwards),
    \item Temporal correlation: Hourly matching (from 1.1.2030 onwards, until 31.12.2029 monthly) and
    \item Geographical correlation
\end{enumerate}
which are required for Power Purchase Agreements (PPA) with RE-installation. Apart from the PPAs, the delegated act also considers the possibility of:
\begin{itemize}
    \item direct connection of RE and electrolysers,
    \item high share of RE in power mix ($>$ 90\%) or the
    \item avoidance of RE curtailment
\end{itemize}
to qualify as "green" hydrogen. These further options are not considered here, since a system-integrated electrolysis in a power system of a (current) share of RE below 90\% is the scope of this study. The sole focus on RE curtailment and hence low total hydrogen volumes is not applicable due to expected hydrogen exports of up to multiple times of Morocco's domestic electricity demand. The green hydrogen definition of the European Commission prior to 1.1.2028 is not applied in this study, since relevant volumes of hydrogen export are expected to materialize from 2028 onwards and is excluded by the scope of this study presented in the \nameref{sec:intro}).

\subsection*{Hydrogen export}
\label{subsec:hydrogen_export}
The amount of hydrogen to be exported or further synthesized for export is  implemented as an exogenous parameter in the range of 0-120~TWh/a, with sensitivity analysis up to 200 TWh/a presented in the \nameref{sec:si} section \nameref{sec:highexportsens}. The system boundary is the country border of Morocco, hence transport options (as shipping or pipeline) are not considered in detail. The profile for hydrogen (or derivatives) export assumed in this study is constant. This represents the operation of pipeline exports as well as the operation of a further hydrogen synthesis with limited flexibility. The export of hydrogen is allowed via a range of ports in Morocco.

The energy system model allows an endogenous spatial export decision, meaning that the total demand (and export profiles) are exogenous, but the model chooses the  cost-optimal export location(s). At each port, the model has the option to build a hydrogen underground or steel tank depending on domestic geological conditions.

\section*{Data availability}
A dataset of the model results is available on \href{https://doi.org/10.5281/zenodo.10951650}{Zenodo} under a CC-BY-4.0 license. 
Technology data was taken from the
\href{https://github.com/pypsa/technology-data}{technology-data repository} (v0.4.0).

\section*{Code availability}
The code to reproduce the experiments is available on \href{https://github.com/energyLS/aldehyde}{GitHub}.
We also refer to the documentations of \href{https://pypsa.readthedocs.io}{PyPSA} and  \href{https://pypsa-earth.readthedocs.io}{PyPSA-Earth}, and the code of 
\href{https://github.com/pypsa-meets-earth/pypsa-earth-sec}{PyPSA-Earth-Sec}.

\addcontentsline{toc}{section}{References}
\renewcommand{\ttdefault}{\sfdefault}
\bibliography{references_mylibrary}

\begin{thebibliography}{10}
\expandafter\ifx\csname url\endcsname\relax
  \def\url#1{\texttt{#1}}\fi
\expandafter\ifx\csname urlprefix\endcsname\relax\def\urlprefix{URL }\fi
\expandafter\ifx\csname href\endcsname\relax
  \def\href#1#2{#2} \def\path#1{#1}\fi

\bibitem{Peters2023}
R.~Peters, J.~Berlekamp, K.~Tockner, C.~Zarfl, {{RePP Africa}} -- a
  georeferenced and curated database on existing and proposed wind, solar, and
  hydropower plants, Scientific Data 10~(1) (Jan. 2023).
\newblock \href {https://doi.org/10.1038/s41597-022-01922-1}
  {\path{doi:10.1038/s41597-022-01922-1}}.

\bibitem{Touili2018}
S.~Touili, A.~A. Merrouni, A.~Azouzoute, Y.~E. Hassouani, A.-i. Amrani, A
  technical and economical assessment of hydrogen production potential from
  solar energy in {{Morocco}}, International Journal of Hydrogen Energy 43~(51)
  (2018) 22777--22796.
\newblock \href {https://doi.org/10.1016/j.ijhydene.2018.10.136}
  {\path{doi:10.1016/j.ijhydene.2018.10.136}}.

\bibitem{Sterl2022}
S.~Sterl, B.~Hussain, A.~Miketa, Y.~Li, B.~Merven, M.~B.~B. Ticha, M.~A.~E.
  Elabbas, W.~Thiery, D.~Russo, An all-{{Africa}} dataset of energy model
  ``supply regions'' for solar photovoltaic and wind power, Scientific Data
  9~(1) (Oct. 2022).
\newblock \href {https://doi.org/10.1038/s41597-022-01786-5}
  {\path{doi:10.1038/s41597-022-01786-5}}.

\bibitem{CAT2021}
{Climate Action Tracker},
  \href{https://climateactiontracker.org/countries/morocco/}{Country summary
  morocco} (2021).
\newline\urlprefix\url{https://climateactiontracker.org/countries/morocco/}

\bibitem{Ersoy2022}
S.~R. Ersoy, J.~{Terrapon-Pfaff}, P.~Viebahn, T.~Pregger, J.~Braun,
  \href{https://wupperinst.org/fa/redaktion/downloads/projects/MENA-Fuels_Teilbericht11_Laenderkurzstudien.pdf}{Synthese
  der {{Kurzstudien}} f{\"u}r {{Jordanien}}, {{Marokko}} und {{Oman}}.
  {{MENA-Fuels}}:{{Teilbericht}} 11}, Tech. rep., Wuppertal Institut, Deutschen
  Zentrum f{\"u}r Luft- und Raumfahrt (DLR) (2022).
\newline\urlprefix\url{https://wupperinst.org/fa/redaktion/downloads/projects/MENA-Fuels_Teilbericht11_Laenderkurzstudien.pdf}

\bibitem{IEA2022}
{International Energy Agency},
  \href{https://www.iea.org/countries/morocco}{Morocco key energy statistics}
  (2022).
\newline\urlprefix\url{https://www.iea.org/countries/morocco}

\bibitem{vanWijk2021}
A.~{van Wijk}, F.~Wouters, Hydrogen--{{The}} bridge between africa and europe,
  in: Shaping an Inclusive Energy Transition, Springer International
  Publishing, 2021, pp. 91--119.
\newblock \href {https://doi.org/10.1007/978-3-030-74586-8_5}
  {\path{doi:10.1007/978-3-030-74586-8_5}}.

\bibitem{AbouSeada2022}
N.~AbouSeada, T.~M. Hatem, Climate action: {{Prospects}} of green hydrogen in
  {{Africa}}, Energy Reports 8 (2022) 3873--3890.
\newblock \href {https://doi.org/10.1016/j.egyr.2022.02.225}
  {\path{doi:10.1016/j.egyr.2022.02.225}}.

\bibitem{vanderZwaan2021}
B.~{van der Zwaan}, S.~Lamboo, F.~D. Longa, Timmermans' dream: {{An}}
  electricity and hydrogen partnership between {{Europe}} and {{North Africa}}
  159 (2021) 112613.
\newblock \href {https://doi.org/10.1016/j.enpol.2021.112613}
  {\path{doi:10.1016/j.enpol.2021.112613}}.

\bibitem{Schellekens2010}
G.~Schellekens, A.~Battaglini, J.~Lilliestam, J.~McDonnell, A.~Patt,
  \href{https://www.pwc.com/gx/en/sustainability/research-insights/assets/renewable-electricity-2050.pdf}{Moving
  towards 100{\textbackslash}\% renewable electricity in {{Europe}}
  {\textbackslash}\& {{North Africa}} by 2050}, Tech. rep. (2010).
\newline\urlprefix\url{https://www.pwc.com/gx/en/sustainability/research-insights/assets/renewable-electricity-2050.pdf}

\bibitem{Cavana2021}
M.~Cavana, P.~Leone, Solar hydrogen from {{North Africa}} to {{Europe}} through
  greenstream: {{A}} simulation-based analysis of blending scenarios and
  production plant sizing, International Journal of Hydrogen Energy 46~(43)
  (2021) 22618--22637.
\newblock \href {https://doi.org/10.1016/j.ijhydene.2021.04.065}
  {\path{doi:10.1016/j.ijhydene.2021.04.065}}.

\bibitem{Touili2022}
S.~Touili, A.~Bouaichi, A.~A. Merrouni, A.-i. Amrani, A.~E. Amrani, Y.~E.
  Hassouani, C.~Messaoudi, Performance analysis and economic competiveness of 3
  different {{PV}} technologies for hydrogen production under the impact of
  arid climatic conditions of {{Morocco}}, International Journal of Hydrogen
  Energy (Aug. 2022).
\newblock \href {https://doi.org/10.1016/j.ijhydene.2022.07.088}
  {\path{doi:10.1016/j.ijhydene.2022.07.088}}.

\bibitem{Timmerberg2019a}
S.~Timmerberg, M.~Kaltschmitt, Hydrogen from renewables: {{Supply}} from
  {{North Africa}} to {{Central Europe}} as blend in existing pipelines --
  {{Potentials}} and costs, Applied Energy 237 (2019) 795--809.
\newblock \href {https://doi.org/10.1016/j.apenergy.2019.01.030}
  {\path{doi:10.1016/j.apenergy.2019.01.030}}.

\bibitem{Sens2022}
L.~Sens, Y.~Piguel, U.~Neuling, S.~Timmerberg, K.~Wilbrand, M.~Kaltschmitt,
  Cost minimized hydrogen from solar and wind -- {{Production}} and supply in
  the {{European}} catchment area, Energy Conversion and Management 265 (2022)
  115742.
\newblock \href {https://doi.org/10.1016/j.enconman.2022.115742}
  {\path{doi:10.1016/j.enconman.2022.115742}}.

\bibitem{Boulakhbar2020}
M.~Boulakhbar, B.~Lebrouhi, T.~Kousksou, S.~Smouh, A.~Jamil, M.~Maaroufi,
  M.~Zazi, Towards a large-scale integration of renewable energies in
  {{Morocco}}, Journal of Energy Storage 32 (2020) 101806.
\newblock \href {https://doi.org/10.1016/j.est.2020.101806}
  {\path{doi:10.1016/j.est.2020.101806}}.

\bibitem{Khouya2020}
A.~Khouya, Levelized costs of energy and hydrogen of wind farms and
  concentrated photovoltaic thermal systems. {{A}} case study in {{Morocco}},
  International Journal of Hydrogen Energy 45~(56) (2020) 31632--31650.
\newblock \href {https://doi.org/10.1016/j.ijhydene.2020.08.240}
  {\path{doi:10.1016/j.ijhydene.2020.08.240}}.

\bibitem{Hampp2023}
J.~Hampp, M.~D{\"u}ren, T.~Brown, Import options for chemical energy carriers
  from renewable sources to {{Germany}}, PLOS ONE 18~(2) (Feb. 2023).
\newblock \href {https://doi.org/10.1371/journal.pone.0281380}
  {\path{doi:10.1371/journal.pone.0281380}}.

\bibitem{Eichhammer2019}
W.~Eichhammer, S.~Oberle, M.~H{\"a}ndel, I.~Boie, T.~Gnann, M.~Wietschel,
  B.~Lux,
  \href{https://publica-rest.fraunhofer.de/server/api/core/bitstreams/5156d310-48ac-4e84-a6ec-3e657b93e198/content}{Study
  on the opportunities of "{{Power-to-X}}" in {{Morocco}}. 10 hypotheses for
  discussion}, Tech. rep., Fraunhofer ISI (2019).
\newline\urlprefix\url{https://publica-rest.fraunhofer.de/server/api/core/bitstreams/5156d310-48ac-4e84-a6ec-3e657b93e198/content}

\bibitem{Zeyen2024}
E.~Zeyen, I.~Riepin, T.~Brown, Temporal regulation of renewable supply for
  electrolytic hydrogen, Environmental Research Letters 19~(2) (2024) 024034.
\newblock \href {https://doi.org/10.1088/1748-9326/ad2239}
  {\path{doi:10.1088/1748-9326/ad2239}}.

\bibitem{Ruhnau2023a}
O.~Ruhnau, J.~Schiele, Flexible green hydrogen: {{The}} effect of relaxing
  simultaneity requirements on project design, economics, and power sector
  emissions, Energy Policy 182 (2023) 113763.
\newblock \href {https://doi.org/10.1016/j.enpol.2023.113763}
  {\path{doi:10.1016/j.enpol.2023.113763}}.

\bibitem{Brauer2022}
J.~Brauer, M.~Villavicencio, J.~Tr{\"u}by,
  \href{https://cadmus.eui.eu/bitstream/handle/1814/74850/RSC_WP_2022_44.pdf}{Green
  hydrogen -- {{How}} grey can it be?}, Tech. rep., European University
  InstituteRobert Schuman Centre for Advanced StudiesThe Florence School of
  Regulation (2022).
\newline\urlprefix\url{https://cadmus.eui.eu/bitstream/handle/1814/74850/RSC_WP_2022_44.pdf}

\bibitem{Giovanniello2024}
M.~A. Giovanniello, A.~N. Cybulsky, T.~Schittekatte, D.~S. Mallapragada,
  \href{https://www.nature.com/articles/s41560-023-01435-0}{The influence of
  additionality and time-matching requirements on the emissions from
  grid-connected hydrogen production}, Nature Energy (2024) 1--11\href
  {https://doi.org/10.1038/s41560-023-01435-0}
  {\path{doi:10.1038/s41560-023-01435-0}}.
\newline\urlprefix\url{https://www.nature.com/articles/s41560-023-01435-0}

\bibitem{Neumann2022}
F.~Neumann, E.~Zeyen, M.~Victoria, T.~Brown, Benefits of a hydrogen network in
  europe, SSRN Electronic Journal (2022).
\newblock \href {https://doi.org/10.2139/ssrn.4173442}
  {\path{doi:10.2139/ssrn.4173442}}.

\bibitem{Oyewo2023}
A.~S. Oyewo, S.~Sterl, S.~Khalili, C.~Breyer, Highly renewable energy systems
  in {{Africa}}: {{Rationale}}, research, and recommendations, Joule 7~(7)
  (2023) 1437--1470.
\newblock \href {https://doi.org/10.1016/j.joule.2023.06.004}
  {\path{doi:10.1016/j.joule.2023.06.004}}.

\bibitem{MarHyStrat2021}
{Royaume du Maroc},
  \href{https://www.mem.gov.ma/Lists/Lst_rapports/Attachments/36/Feuille%20de%20route%20de%20hydrog%C3%A8ne%20vert.pdf}{Feuille
  de route de hydrog{\`e}ne vert}, Tech. rep. (2021).
\newline\urlprefix\url{https://www.mem.gov.ma/Lists/Lst_rapports/Attachments/36/Feuille%20de%20route%20de%20hydrog%C3%A8ne%20vert.pdf}

\bibitem{Terrapon-Pfaff2019}
J.~{Terrapon-Pfaff}, T.~Fink, P.~Viebahn, E.~M. Jamea, Social impacts of
  large-scale solar thermal power plants: {{Assessment}} results for the
  {{NOORO I}} power plant in {{Morocco}}, Renewable and Sustainable Energy
  Reviews 113 (2019) 109259.
\newblock \href {https://doi.org/10.1016/j.rser.2019.109259}
  {\path{doi:10.1016/j.rser.2019.109259}}.

\bibitem{Hanger2016}
S.~Hanger, N.~Komendantova, B.~Schinke, D.~Zejli, A.~Ihlal, A.~Patt, Community
  acceptance of large-scale solar energy installations in developing countries:
  {{Evidence}} from {{Morocco}}, Energy Research \& Social Science 14 (2016)
  80--89.
\newblock \href {https://doi.org/10.1016/j.erss.2016.01.010}
  {\path{doi:10.1016/j.erss.2016.01.010}}.

\bibitem{Merten2023}
F.~Merten, A.~Scholz,
  \href{https://epub.wupperinst.org/frontdoor/deliver/index/docId/8344/file/8344_Wasserstoffkosten.pdf}{{Metaanalyse
  zu Wasserstoffkosten und -bedarfen f{\"u}r die CO2-neutrale Transformation}},
  Tech. rep.
\newline\urlprefix\url{https://epub.wupperinst.org/frontdoor/deliver/index/docId/8344/file/8344_Wasserstoffkosten.pdf}

\bibitem{Tries2023b}
C.~Tries, F.~Hofmann, T.~Brown, Benefits from {{Islanding Green Hydrogen
  Production}} (Oct. 2023).
\newblock \href {http://arxiv.org/abs/2310.12606} {\path{arXiv:2310.12606}},
  \href {https://doi.org/10.48550/arXiv.2310.12606}
  {\path{doi:10.48550/arXiv.2310.12606}}.

\bibitem{Galimova2023}
T.~Galimova, M.~Fasihi, D.~Bogdanov, C.~Breyer, Impact of international
  transportation chains on cost of green e-hydrogen: {{Global}} cost of
  hydrogen and consequences for {{Germany}} and {{Finland}}, Applied Energy 347
  (2023) 121369.
\newblock \href {https://doi.org/10.1016/j.apenergy.2023.121369}
  {\path{doi:10.1016/j.apenergy.2023.121369}}.

\bibitem{Verpoort2023}
P.~C. Verpoort, L.~Gast, A.~Hofmann, F.~Ueckerdt, Estimating the renewables
  pull in future global green value chains (Apr. 2023).
\newblock \href {https://doi.org/10.21203/rs.3.rs-2743794/v1}
  {\path{doi:10.21203/rs.3.rs-2743794/v1}}.

\bibitem{DEA2019TechnologyData}
{Danish Energy Agency (DEA)},
  \href{https://ens.dk/en/our-services/projections-and-models/technology-data}{Technology
  data} (2019).
\newline\urlprefix\url{https://ens.dk/en/our-services/projections-and-models/technology-data}

\bibitem{Zeyen2023}
E.~Zeyen, M.~Victoria, T.~Brown, Endogenous learning for green hydrogen in a
  sector-coupled energy model for {{Europe}}, Nature Communications 14~(1)
  (Jun. 2023).
\newblock \href {https://doi.org/10.1038/s41467-023-39397-2}
  {\path{doi:10.1038/s41467-023-39397-2}}.

\bibitem{Wang2023}
S.~Wang, Z.~Hausfather, S.~Davis, J.~Lloyd, E.~B. Olson, L.~Liebermann, G.~D.
  {N{\'u}{\~n}ez-Mujica}, J.~McBride, Future demand for electricity generation
  materials under different climate mitigation scenarios, Joule 7~(2) (2023)
  309--332.
\newblock \href {https://doi.org/10.1016/j.joule.2023.01.001}
  {\path{doi:10.1016/j.joule.2023.01.001}}.

\bibitem{Caldera2016}
U.~Caldera, D.~Bogdanov, C.~Breyer, Local cost of seawater {{RO}} desalination
  based on solar {{PV}} and wind energy: {{A}} global estimate, Desalination
  385 (2016) 207--216.
\newblock \href {https://doi.org/10.1016/j.desal.2016.02.004}
  {\path{doi:10.1016/j.desal.2016.02.004}}.

\bibitem{Parzen2023}
M.~Parzen, H.~{Abdel-Khalek}, E.~Fedotova, M.~Mahmood, M.~M. Frysztacki,
  J.~Hampp, L.~Franken, L.~Schumm, F.~Neumann, D.~Poli, A.~Kiprakis,
  D.~Fioriti, {{PyPSA-Earth}}. {{A}} new global open energy system optimization
  model demonstrated in {{Africa}}, Applied Energy 341 (2023) 121096.
\newblock \href {https://doi.org/10.1016/j.apenergy.2023.121096}
  {\path{doi:10.1016/j.apenergy.2023.121096}}.

\bibitem{Abdel-Khalek2024}
H.~{Abdel-Khalek}, L.~Schumm, E.~Jalbout, M.~Parzen, C.~Schau{\ss}, D.~Fioriti,
  \href{https://papers.ssrn.com/abstract=4743242}{Sector-{{Coupled
  Pypsa-Earth}}, a {{Global Open-Source Multi-Energy System Model}}} (Feb.
  2024).
\newblock \href {https://doi.org/10.2139/ssrn.4743242}
  {\path{doi:10.2139/ssrn.4743242}}.
\newline\urlprefix\url{https://papers.ssrn.com/abstract=4743242}

\bibitem{unstats2023}
{United Nations Statistics Division},
  \href{https://unstats.un.org/unsd/energystats/}{Energy {{Statistics}}}
  (2023).
\newline\urlprefix\url{https://unstats.un.org/unsd/energystats/}

\bibitem{Muller2023}
V.~P. M{\"u}ller, N.~Pieton, V.~Leninova, H.~{Abdel-Khalek}, M.~F.~A. Sinha,
  National energy demand projections using {{LEAP}} and {{Excel}} (2023).
\newblock \href {https://doi.org/10.5281/ZENODO.8153736}
  {\path{doi:10.5281/ZENODO.8153736}}.

\bibitem{Brown2018a}
T.~Brown, J.~H{\"o}rsch, D.~Schlachtberger, {{PyPSA}}: {{Python}} for power
  system analysis, Journal of Open Research Software 6 (2018).
\newblock \href {https://doi.org/10.5334/jors.188}
  {\path{doi:10.5334/jors.188}}.

\bibitem{Rim2021}
B.~Rim, C.~Abdelilah, D.~Atar, H.~Ibtissem, M.~Mariano, N.~Yassine,
  \href{https://www.policycenter.ma/sites/default/files/2022-11/PB-26-21-Enel-Green-Power-EGP-EN-PART%20IV_0.pdf}{Morocco's
  {{Decarbonization Pathway}} - {{Part IV}}: {{Policy Recommmendations}}},
  Tech. rep. (Jul. 2021).
\newline\urlprefix\url{https://www.policycenter.ma/sites/default/files/2022-11/PB-26-21-Enel-Green-Power-EGP-EN-PART%20IV_0.pdf}

\bibitem{Hofmann2021}
F.~Hofmann, J.~Hampp, F.~Neumann, T.~Brown, J.~H{\"o}rsch, Atlite: {{A}}
  lightweight python package for calculating renewable power potentials and
  time series, Journal of Open Source Software 6~(62) (2021) 3294.
\newblock \href {https://doi.org/10.21105/joss.03294}
  {\path{doi:10.21105/joss.03294}}.

\bibitem{Hersbach2020}
H.~Hersbach, B.~Bell, P.~Berrisford, S.~Hirahara, A.~Hor{\'a}nyi,
  J.~{Mu{\~n}oz-Sabater}, J.~Nicolas, C.~Peubey, R.~Radu, D.~Schepers,
  A.~Simmons, C.~Soci, S.~Abdalla, X.~Abellan, G.~Balsamo, P.~Bechtold,
  G.~Biavati, J.~Bidlot, M.~Bonavita, G.~Chiara, P.~Dahlgren, D.~Dee,
  M.~Diamantakis, R.~Dragani, J.~Flemming, R.~Forbes, M.~Fuentes, A.~Geer,
  L.~Haimberger, S.~Healy, R.~J. Hogan, E.~H{\'o}lm, M.~Janiskov{\'a},
  S.~Keeley, P.~Laloyaux, P.~Lopez, C.~Lupu, G.~Radnoti, P.~Rosnay, I.~Rozum,
  F.~Vamborg, S.~Villaume, J.-N. Th{\'e}paut, The {{ERA5}} global reanalysis,
  Quarterly Journal of the Royal Meteorological Society 146~(730) (2020)
  1999--2049.
\newblock \href {https://doi.org/10.1002/qj.3803} {\path{doi:10.1002/qj.3803}}.

\bibitem{Pfeifroth2017}
U.~Pfeifroth, S.~Kothe, R.~M{\"u}ller, J.~Trentmann, R.~Hollmann, P.~Fuchs,
  M.~Werscheck, Surface radiation data set - heliosat ({{SARAH}}) - edition 2
  (2017).
\newblock \href {https://doi.org/10.5676/EUM_SAF_CM/SARAH/V002}
  {\path{doi:10.5676/EUM_SAF_CM/SARAH/V002}}.

\bibitem{Buchhorn2020}
M.~Buchhorn, B.~Smets, L.~Bertels, B.~D. Roo, M.~Lesiv, N.-E. Tsendbazar,
  M.~Herold, S.~Fritz, Copernicus global land service: {{Land}} cover 100m:
  Collection 3: Epoch 2018: {{Globe}} (2020).
\newblock \href {https://doi.org/10.5281/ZENODO.3518038}
  {\path{doi:10.5281/ZENODO.3518038}}.

\bibitem{Parzen2022}
M.~Parzen, H.~{Abdel-Khalek}, E.~Fedorova, M.~Mahmood, M.~M. Frysztacki,
  J.~Hampp, L.~Franken, L.~Schumm, F.~Neumann, D.~Poli, A.~Kiprakis,
  D.~Fioriti, {{PyPSA-Earth}}. {{A}} new global open energy system optimization
  model demonstrated in africa (2022).
\newblock \href {https://doi.org/10.48550/ARXIV.2209.04663}
  {\path{doi:10.48550/ARXIV.2209.04663}}.

\bibitem{Horsch2018}
J.~H{\"o}rsch, F.~Hofmann, D.~Schlachtberger, T.~Brown, {{PyPSA-Eur}}: {{An}}
  open optimisation model of the {{European}} transmission system, Energy
  Strategy Reviews 22 (2018) 207--215.
\newblock \href {https://doi.org/10.1016/j.esr.2018.08.012}
  {\path{doi:10.1016/j.esr.2018.08.012}}.

\bibitem{Powerplantmatching2019}
{Powerplantmatching},
  \href{https://github.com/PyPSA/powerplantmatching}{Powerplantmatching tool}
  (2023).
\newline\urlprefix\url{https://github.com/PyPSA/powerplantmatching}

\bibitem{IRENA2022}
{International Renewable Energy Agency},
  \href{https://pxweb.irena.org/pxweb/en/IRENASTAT/}{{{IRENASTAT Online Data
  Query Tool}}} (2023).
\newline\urlprefix\url{https://pxweb.irena.org/pxweb/en/IRENASTAT/}

\bibitem{OpenStreetMap2022}
{OpenStreetMap contributors},
  \href{https://www.openstreetmap.org}{{{OpenStreetMap}}} (2022).
\newline\urlprefix\url{https://www.openstreetmap.org}

\bibitem{GlobalEnergyMonitor}
G.~E. Monitor, \href{https://www.gem.wiki/Main_Page}{Global energy monitor
  wiki} (2024).
\newline\urlprefix\url{https://www.gem.wiki/Main_Page}

\bibitem{Rachidi2022}
I.~Rachidi, \href{https://carnegieendowment.org/sada/87055}{Morocco and
  {{Algeria}}: {{A Long Rivalry}}} (2022).
\newline\urlprefix\url{https://carnegieendowment.org/sada/87055}

\bibitem{GEM2023b}
G.~E. Monitor,
  \href{https://www.gem.wiki/Nigeria-Morocco_Gas_Pipeline}{Nigeria-{{Morocco
  Gas Pipeline}}} (2023).
\newline\urlprefix\url{https://www.gem.wiki/Nigeria-Morocco_Gas_Pipeline}

\bibitem{Tong2020}
W.~Tong, M.~Forster, F.~Dionigi, S.~Dresp, R.~Sadeghi~Erami, P.~Strasser, A.~J.
  Cowan, P.~Farr{\`a}s,
  \href{https://www.nature.com/articles/s41560-020-0550-8}{Electrolysis of
  low-grade and saline surface water}, Nature Energy 5~(5) (2020) 367--377.
\newblock \href {https://doi.org/10.1038/s41560-020-0550-8}
  {\path{doi:10.1038/s41560-020-0550-8}}.
\newline\urlprefix\url{https://www.nature.com/articles/s41560-020-0550-8}

\bibitem{Maddocks2015}
A.~Maddocks, R.~S. Young, P.~Reig,
  \href{https://www.wri.org/insights/ranking-worlds-most-water-stressed-countries-2040}{Ranking
  the {{World}}'s {{Most Water-Stressed Countries}} in 2040} (Wed, 08/26/2015 -
  00:01).
\newline\urlprefix\url{https://www.wri.org/insights/ranking-worlds-most-water-stressed-countries-2040}

\bibitem{Caldera2020}
U.~Caldera, C.~Breyer, Strengthening the global water supply through a
  decarbonised global desalination sector and improved irrigation systems,
  Energy 200 (2020) 117507.
\newblock \href {https://doi.org/10.1016/j.energy.2020.117507}
  {\path{doi:10.1016/j.energy.2020.117507}}.

\bibitem{Kettani2020}
M.~Kettani, P.~Bandelier, Techno-economic assessment of solar energy coupling
  with large-scale desalination plant: {{The}} case of {{Morocco}},
  Desalination 494 (2020) 114627.
\newblock \href {https://doi.org/10.1016/j.desal.2020.114627}
  {\path{doi:10.1016/j.desal.2020.114627}}.

\bibitem{Thomann2022}
J.~Thomann, F.~{Marscheider-Weidemann}, A.~Stamm, L.~Lorych, C.~Hank, F.~Weise,
  L.~Edenhofer, Z.~Thiel,
  \href{https://publica-rest.fraunhofer.de/server/api/core/bitstreams/e974188e-80dd-49b2-984f-532f39620012/content}{{{HYPAT
  Working Paper}} 01/2022. {{Background}} paper on sustainable green hydrogen
  and synthesis products}, Tech. rep. (2022).
\newline\urlprefix\url{https://publica-rest.fraunhofer.de/server/api/core/bitstreams/e974188e-80dd-49b2-984f-532f39620012/content}

\bibitem{Dresp2019}
S.~Dresp, F.~Dionigi, M.~Klingenhof, P.~Strasser, Direct {{Electrolytic
  Splitting}} of {{Seawater}}: {{Opportunities}} and {{Challenges}}, ACS Energy
  Letters 4~(4) (2019) 933--942.
\newblock \href {https://doi.org/10.1021/acsenergylett.9b00220}
  {\path{doi:10.1021/acsenergylett.9b00220}}.

\bibitem{Tonelli2023}
D.~Tonelli, L.~Rosa, P.~Gabrielli, K.~Caldeira, A.~Parente, F.~Contino, Global
  land and water limits to electrolytic hydrogen production using wind and
  solar resources, Nature Communications 14~(1) (2023) 5532.
\newblock \href {https://doi.org/10.1038/s41467-023-41107-x}
  {\path{doi:10.1038/s41467-023-41107-x}}.

\bibitem{Ruhnau2022}
O.~Ruhnau, A.~Eicke, R.~Sgarlato, T.~Tr{\"o}ndle, L.~Hirth, Cost-potential
  curves of onshore wind energy: The role of disamenity costs, Environmental
  and Resource Economics (Nov. 2022).
\newblock \href {https://doi.org/10.1007/s10640-022-00746-2}
  {\path{doi:10.1007/s10640-022-00746-2}}.

\bibitem{Commission2023}
E.~Commission,
  \href{https://energy.ec.europa.eu/publications/delegated-regulation-union-methodology-rfnbos_en}{Delegated
  regulation on {{Union}} methodology for {{RFNBOs}}}, Tech. rep. (2023).
\newline\urlprefix\url{https://energy.ec.europa.eu/publications/delegated-regulation-union-methodology-rfnbos_en}

\end{thebibliography}

\section*{Acknowledgements}

We are grateful for helpful comments by Alexander Meisinger and Anton Achhammer.

We gratefully acknowledge funding from the H2Global meets Africa project (03SF0703A) and the HYPAT project (03SF0620A) by the German Federal Ministry of Education and Research.

\section*{Author Contributions}

\textbf{L.S.}:
Conceptualization,
Data curation,
Formal Analysis,
Funding acquisition,
Investigation,
Methodology,
Project administration,
Software,
Validation,
Visualization,
Writing - original draft,
Writing - review \& editing;
\textbf{H.A-K.}:
Data curation,
Methodology,
Project administration,
Software,
Validation,
Writing - review \& editing;
\textbf{T.B.}:
Conceptualization,
Formal Analysis,
Methodology,
Supervision,
Writing - review \& editing;
\textbf{F.U.}:
Conceptualization,
Formal Analysis,
Supervision,
Visualization,
Writing - review \& editing;
\textbf{M.S.}:
Funding acquisition,
Project administration,
Resources,
Supervision,
Writing - review \& editing;
\textbf{M.P.}:
Methodology,
Software,
Writing - review \& editing;
\textbf{D.F.}:
Conceptualization,
Formal Analysis,
Methodology,
Software,
Writing - review \& editing.

\section*{Declaration of Interests}

The authors declare no competing interests.

\section*{Materials \& Correspondence}

Requests for further information, resources and materials should be directed to and will be fulfilled by the lead contact, Leon Schumm
(\href{mailto:leon1.schumm@oth-regensburg.de}{leon1.schumm@oth-regensburg.de}).

\newpage
\onecolumn

\makeatletter
\renewcommand \thesection{S\@arabic\c@section}
\renewcommand\thetable{S\@arabic\c@table}
\renewcommand \thefigure{S\@arabic\c@figure}
\makeatother
\renewcommand{\citenumfont}[1]{S#1}
\setcounter{equation}{0}
\setcounter{figure}{0}
\setcounter{table}{0}
\setcounter{section}{0}

\section*{Supplementary Information}
\label{sec:si}

\section{Energy demand}

\subsection{Diffusion of Battery Electric Vehicles}
\label{subsec:bev_diffusion}
\begin{figure*}[h]
    \centering
    \includegraphics[width=0.7\linewidth]{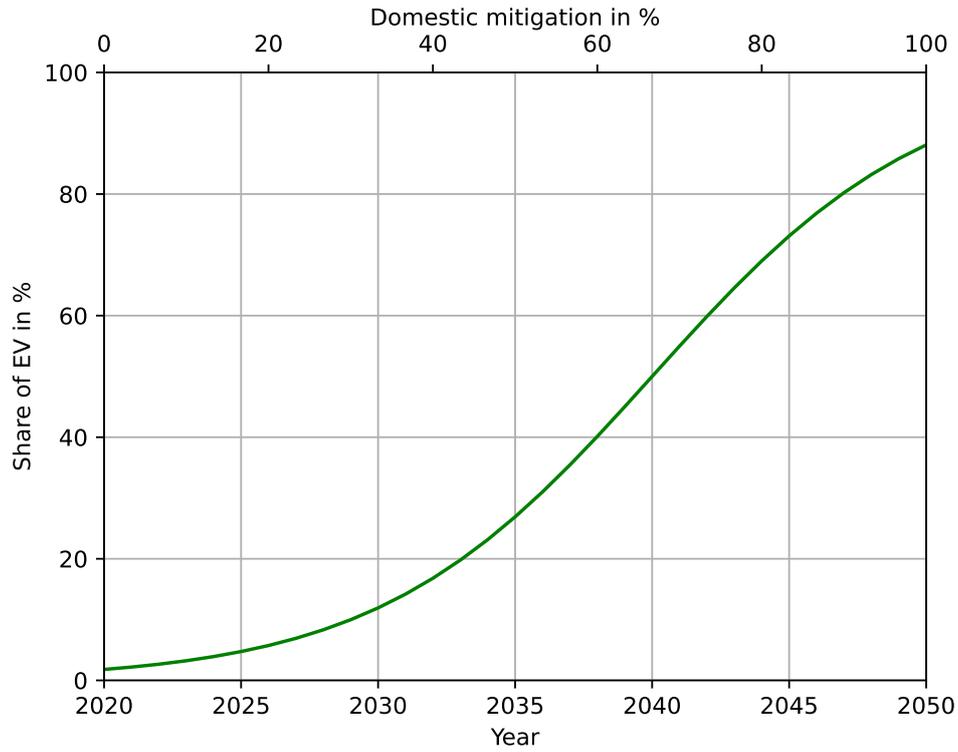}
    \caption{Market diffusion of Battery Electricity Vehicles in Morocco, synthesized based on an s-curve with a growth rate $k=0.2$ and inflection point $x_0=2040$.}
    \label{fig:bev_diffusion}
\end{figure*}

Figure \ref{fig:bev_diffusion} shows the diffusion of battery electric vehicles depending on the domestic climate change mitigation, displaying electric vehicle stock in Morocco. The share rises from 2\% (today's levels) up to a share of 88\% at 100\% domestic climate change mitigation. %

\clearpage

\clearpage

\section{Electricity supply}

\begin{figure}[h]
    \centering
    \includegraphics[width=0.9\linewidth]{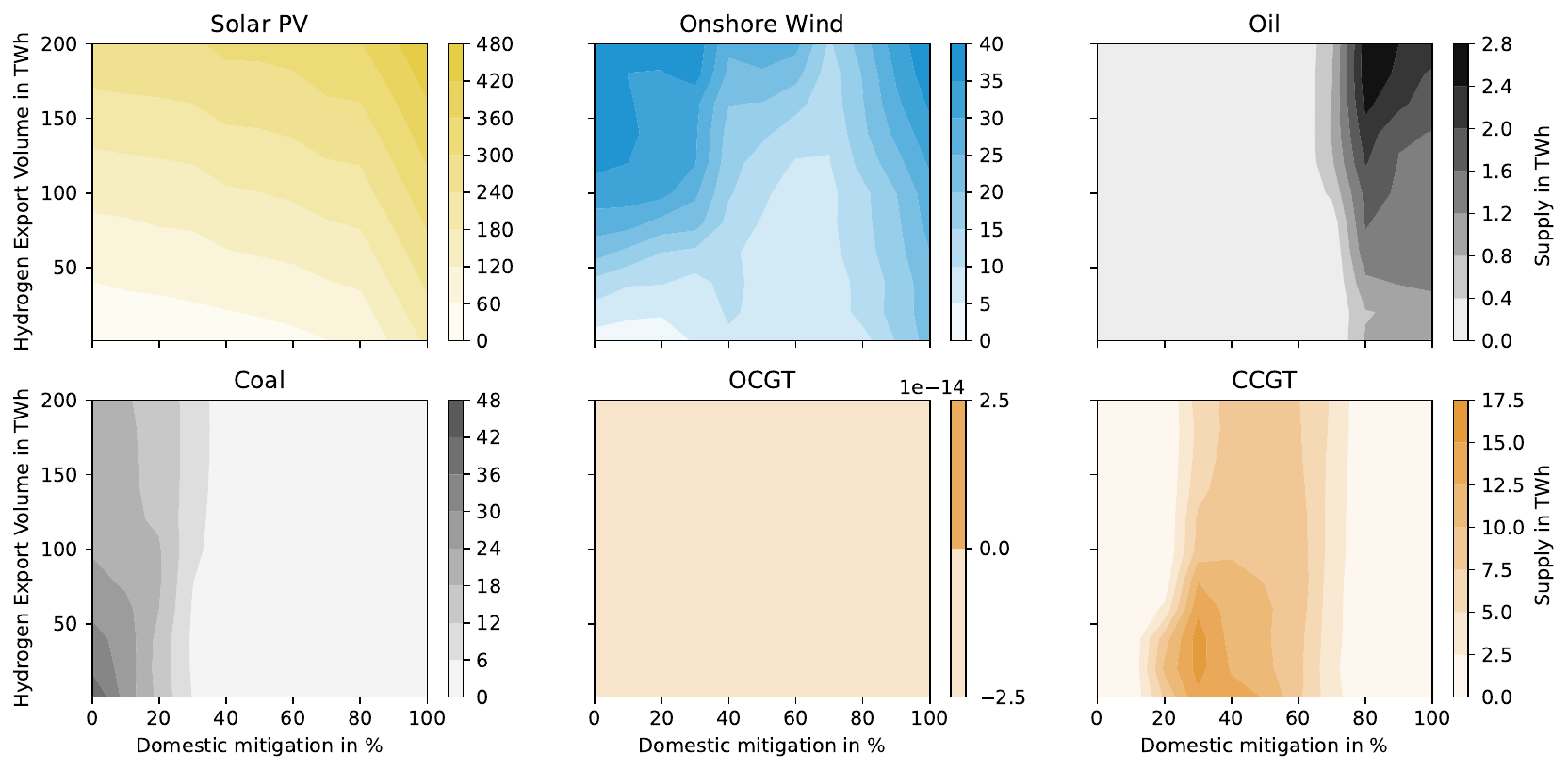}
    \caption{Electricity supply}
    \label{fig:supply}
\end{figure}

\begin{figure}[h]
    \centering
    \includegraphics[width=0.9\linewidth]{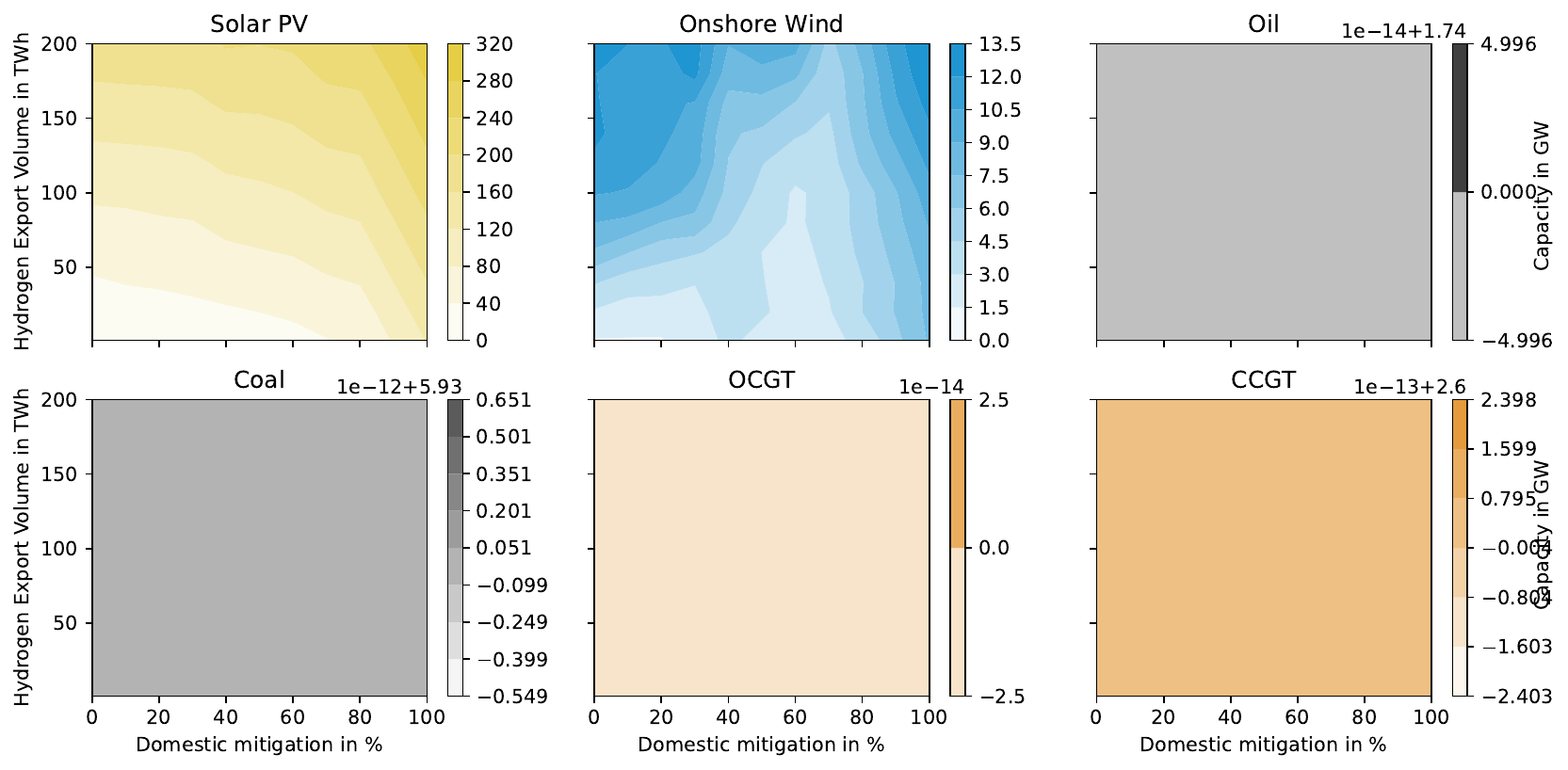}
    \caption{Electricity capacities}
    \label{fig:el-cap}
\end{figure}

\begin{figure}[h]
    \centering
    \includegraphics[width=0.9\linewidth]{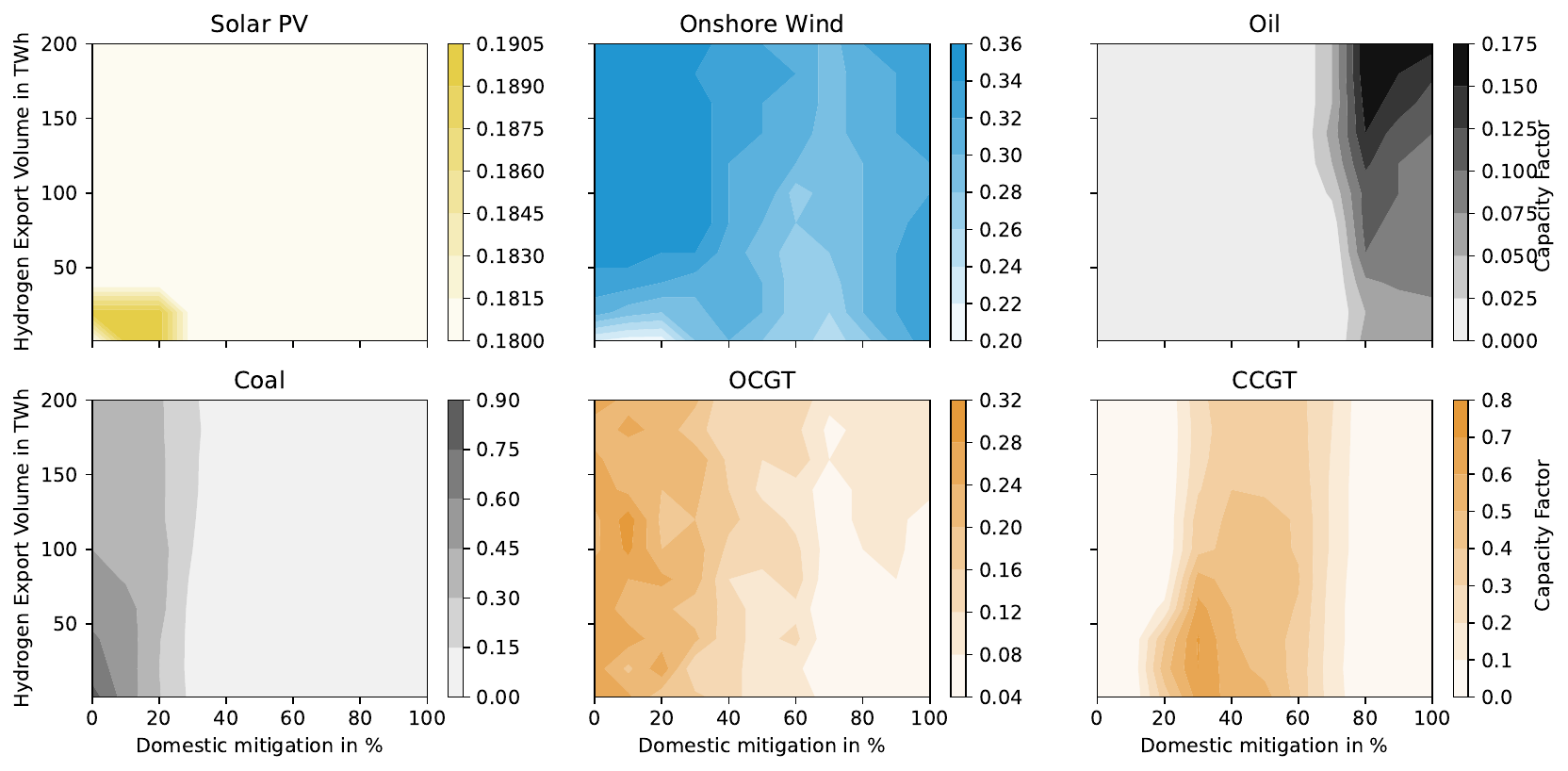}
    \caption{Electricity capacity factors}
    \label{fig:el_cf}
\end{figure}

The electricity supply with hourly temporal matching is displayed in Figure \ref{fig:supply}, along with the generation capacities
in Figure \ref{fig:el-cap} and capacity factors in Figure  \ref{fig:el_cf}.

\subsection{Electricity supply}
\label{subsec:el-supply}

\subsection{Electricity capacities}
\label{subsec:el-cap}

\subsection{Electricity capacity factors}
\label{subsec:el-cf}

\clearpage
\begin{figure}
    \centering
        \begin{subfigure}[h]{0.49\textwidth}
            \centering
        \includegraphics[width=\textwidth]{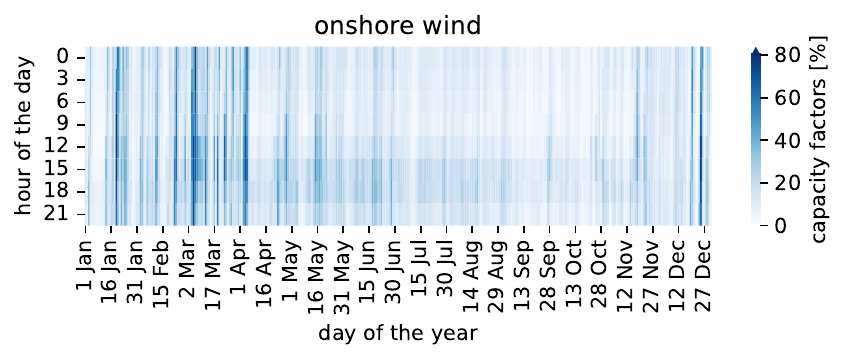}
    \end{subfigure}
    \begin{subfigure}[h]{0.49\textwidth}
        \centering
        \includegraphics[width=\textwidth]{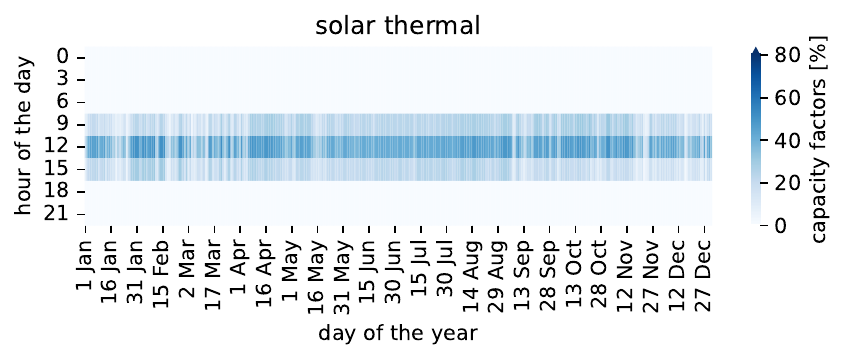}
    \end{subfigure}
    \begin{subfigure}[h]{0.49\textwidth}
        \centering
        \includegraphics[width=\textwidth]{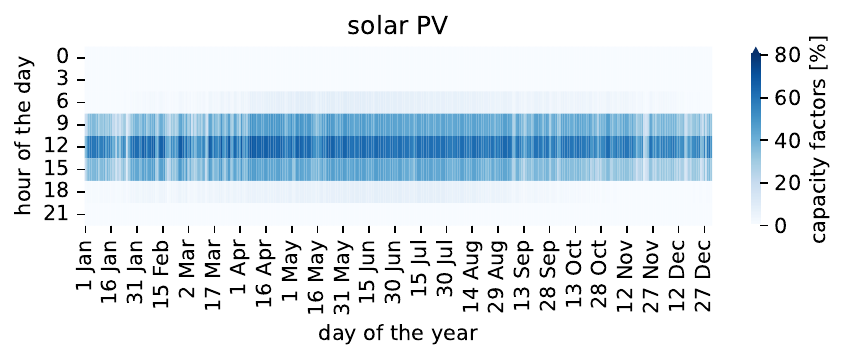}
    \end{subfigure}
    \begin{subfigure}[h]{0.49\textwidth}
        \centering
        \includegraphics[width=\textwidth]{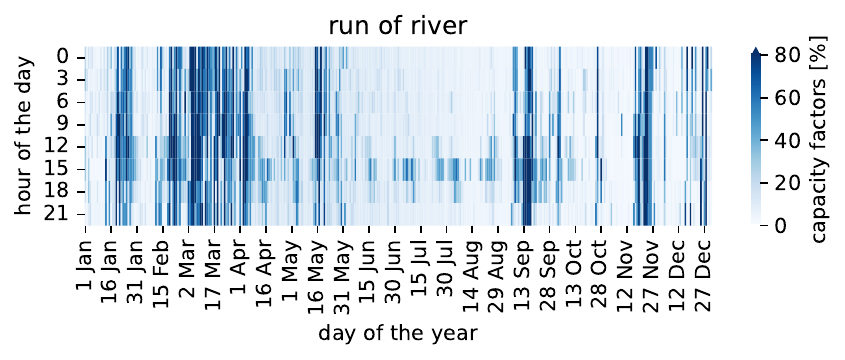}
    \end{subfigure}
    \begin{subfigure}[h]{0.49\textwidth}
        \centering
        \includegraphics[width=\textwidth]{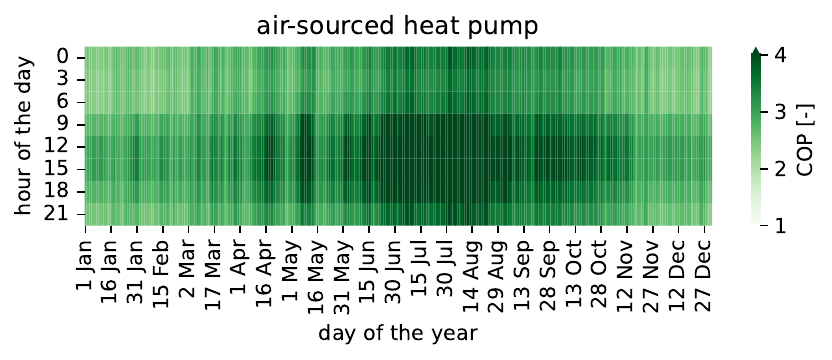}
    \end{subfigure}
    \begin{subfigure}[h]{0.49\textwidth}
        \centering
        \includegraphics[width=\textwidth]{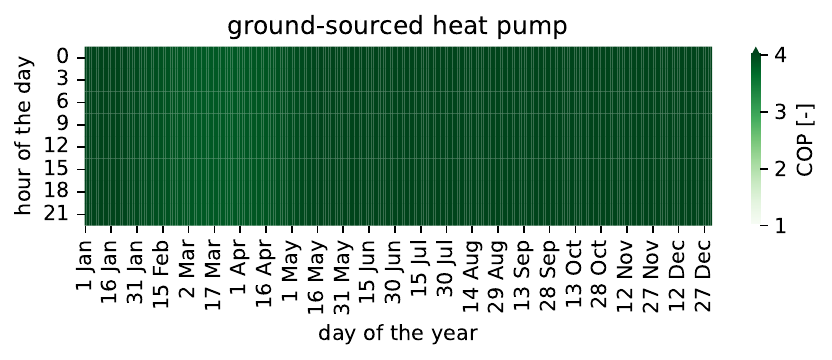}
    \end{subfigure}
    \caption{Capacity factors of renewable sources, graph style adapted from Neumann et al. \cite{Neumann2022}}
    \label{fig:ren-cfs}
\end{figure}

\subsection{Electrolyser and Fischer-Tropsch capacity}

\begin{figure*}[h] %
    \centering
    \begin{subfigure}[b]{0.45\linewidth}
        \centering
        \includegraphics[width=\linewidth]{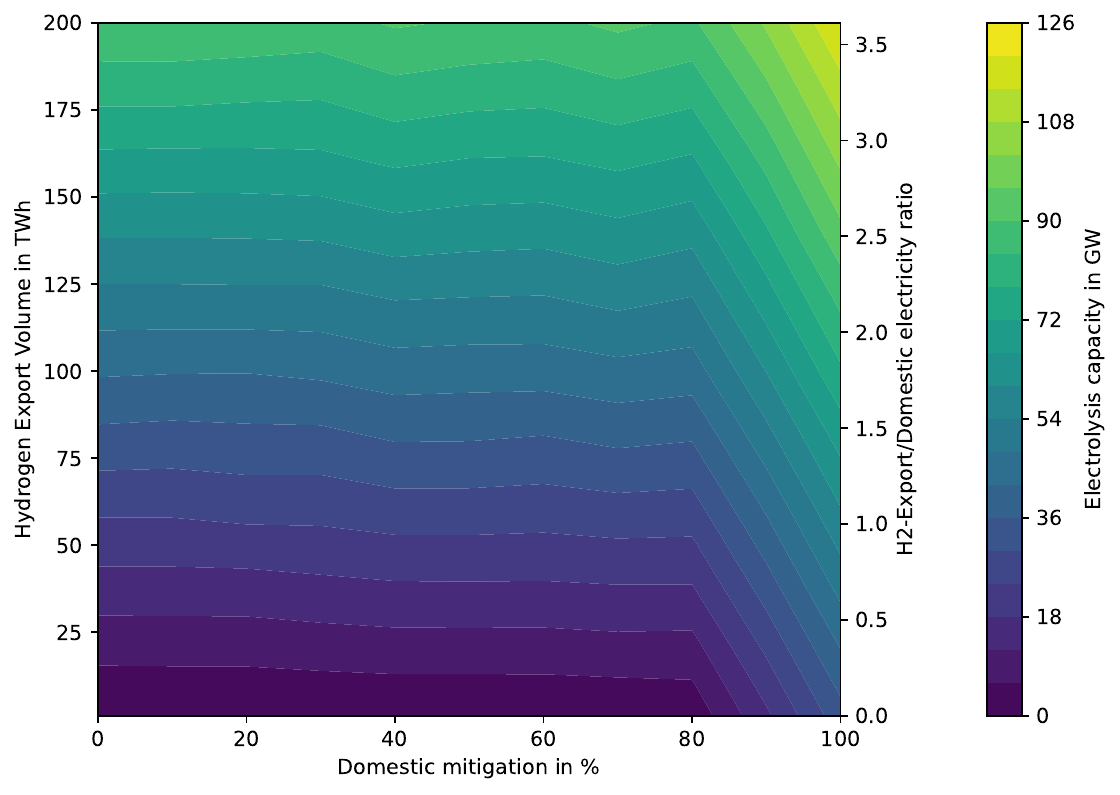}
        \caption{Electrolysis optimal capacity}
        \label{fig:ely-p-nom-opt}
    \end{subfigure}
    \hfill
    \begin{subfigure}[b]{0.45\linewidth}
        \centering
        \includegraphics[width=\linewidth]{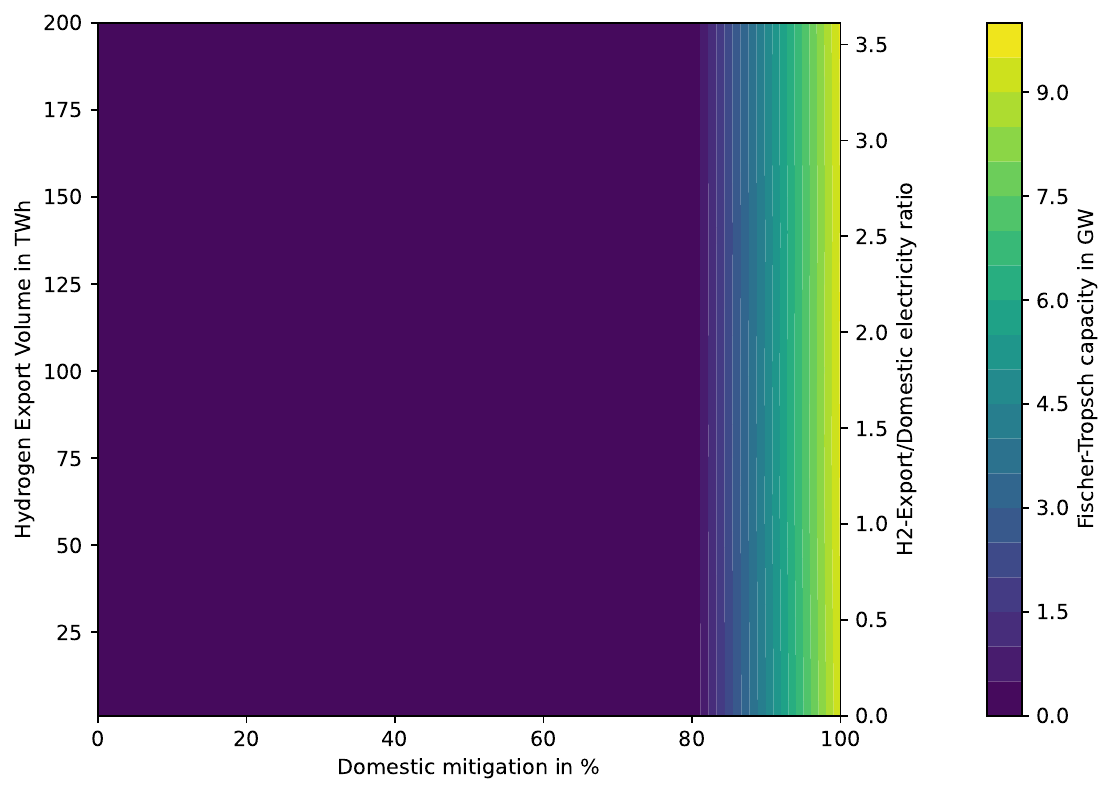}
        \caption{Fischer-Tropsch optimal capacity}
        \label{fig:ft-p-nom-opt}
    \end{subfigure}

    \hfill

    \caption{Electrolysis and Fischer-Tropsch capacities}
    \label{fig:ely-ft-p-nom-opt}
\end{figure*}

\clearpage
\section{Energy balances}

\subsection{Oil balance}
\label{subsec:oil-balance}
Figure \ref{fig:oil-balance} shows the oil balance at 1 TWh/a export at increasing domestic climate change mitigation.

\begin{figure*}[h]
    \centering
    \includegraphics[trim={0cm 0cm 0cm 1cm}, clip, width=0.5\linewidth]{../workflow/subworkflows/pypsa-earth-sec/results/\runstandard/0exp-only/graphs/balances-oil.pdf}
    \caption{Oil balance at 1 TWh/a export at increasing domestic climate change mitigation. The exogenously defined oil demand for land transport substantially decreases due to electrification, whereas oil demands for industrial naphta, aviation, agriculture are constant. The demand is met by fossil oil up to 80\% climate change mitigation, then gradually replaced by Fischer-Tropsch fuels.}
    \label{fig:oil-balance}
\end{figure*}

\clearpage

\section{Temporal hydrogen regulation}
\label{subsec:gh_constraint_effects}

\subsection{Hourly matching prevents additional emissions induced by hydrogen exports}
Figure \ref{fig:barplotscons} displays the electricity balance ramp up from 1--120 TWh/a at 0\% domestic climate change mitigation with (\ref{fig:balances_AC_monthlymatch}) and without (\ref{fig:balances_AC_nogreen}) temporal hydrogen regulation.

\begin{figure*}[h!]
    \centering
    \begin{subfigure}[b]{0.49\linewidth}
        \centering
        \includegraphics[trim={0cm 0cm 0cm 1cm}, clip, width=\linewidth]{../workflow/subworkflows/pypsa-earth-sec/results/\runstandard/co2l20-only/120/graphs/balances-AC.pdf}
        \caption{Electricity balance {\bf with} temporal hydrogen regulation}
        \label{fig:balances_AC_monthlymatch}
    \end{subfigure}
    \hfill
    \begin{subfigure}[b]{0.49\linewidth}
        \centering
        \includegraphics[trim={0cm 0cm 0cm 1cm}, clip, width=\linewidth]{../workflow/subworkflows/pypsa-earth-sec/results/\runsensnogreenhy/co2l20-only/120/graphs/balances-AC.pdf}
        \caption{Electricity balance {\bf without} temporal hydrogen regulation}
        \label{fig:balances_AC_nogreen}
    \end{subfigure}
    \hfill
    \caption{Electricity balance ramp up from 1--120 TWh/a at 0\% domestic climate change mitigation with (\ref{fig:balances_AC_monthlymatch}) and without (\ref{fig:balances_AC_nogreen}) temporal hydrogen regulation}
    \label{fig:barplotscons}
\end{figure*}

\subsection{Effects of temporal matching in a high export and low climate change mitigation scenario on total system costs}

\begin{figure*}[h]
    \centering
    \includegraphics[trim={0cm 0cm 0cm 0.65cm}, clip, width=0.6\linewidth]{../results/gh/abs_sin_tech_120_Co2L2.0.pdf}
    \caption{Total system costs at 120 TWh/a export and 0\% domestic climate change mitigation. Stricter temporal hydrogen regulation mainly increases the total CAPEX of additional solar PV, electrolysis and hydrogen storage. In return, the OPEX of fossil generation (mainly gas), decreases. The large share of oil OPEX is independent of temporal hydrogen regulation, since these costs are mainly linked to combustion engine cars with demands independent of temporal hydrogen regulation.}
    \label{fig:tsc-120-0}
\end{figure*}

\clearpage
\section{Hydrogen cost breakdown}
\label{subsec:electrolysis_op_comp}

The price of hydrogen depends on single cost components as well as the temporal hydrogen regulation. Figure \ref{fig:costbreakdown_hourly} displays the cost breakdown of hydrogen 
whereas the combined cost in Figure \ref{fig:hydrogen-cost-hourly} differs from the price of hydrogen in Figure \ref{fig:hydrogen-price-hourly}. This is due to the temporal hydrogen regulation, the costs for the required installation of additional RE capacities are not reflected in the cost breakdown but included in the price of hydrogen. This effect is most striking in high export and low domestic climate change mitigation scenarios. 
In constrast, if there is no temporal hydrogen regulation applied, both hydrogen cost (s. Fig. \ref{fig:hydrogen-cost-norule}) and hydrogen price (s. Fig. \ref{fig:hydrogen-price-norule}) show similar trends across all domestic climate change mitigation and export scenarios.

\begin{figure*}[h!]
    \centering
    \begin{subfigure}[b]{0.45\linewidth}
        \centering
        \includegraphics[width=\linewidth]{graphics/integrated_comp/contour_capex_ely_rel_20_filterTrue_nFalse_exp200.pdf}
        \caption{CAPEX share of hydrogen costs}
        \label{fig:capex-rel-hourly}
    \end{subfigure}
    \hfill
    \begin{subfigure}[b]{0.45\linewidth}
        \centering
        \includegraphics[width=\linewidth]{graphics/integrated_comp/contour_cf_electrolysis_20_filterTrue_nFalse_exp200.pdf}
        \caption{Electrolysis capacity factor}
        \label{fig:ely-cf-hourly}
    \end{subfigure}
    \hfill
    \begin{subfigure}[b]{0.45\linewidth}
        \centering
        \includegraphics[width=\linewidth]{graphics/integrated_comp/contour_opex_ely_rel_20_filterTrue_nFalse_exp200.pdf}
        \caption{OPEX share of hydrogen costs}
        \label{fig:opex-rel-hourly}
    \end{subfigure}
    \hfill
    \begin{subfigure}[b]{0.45\linewidth}
        \centering
        \includegraphics[width=\linewidth]{graphics/integrated_comp/contour_mg_AC_inclu_H2_El_all_20_filterTrue_nFalse_exp200.pdf}
        \caption{Price of electricity (electrolysis demand weighted)}
        \label{fig:electricity-price-hourly}
    \end{subfigure}
    \hfill
    \begin{subfigure}[b]{0.45\linewidth}
        \centering
        \includegraphics[width=\linewidth]{graphics/integrated_comp/contour_lcoh_compo_20_filterTrue_nFalse_exp200.pdf}
        \caption{Total hydrogen cost}
        \label{fig:hydrogen-cost-hourly}
    \end{subfigure}
    \hfill
    \begin{subfigure}[b]{0.45\linewidth}
        \centering
        \includegraphics[width=\linewidth]{graphics/integrated_comp/contour_mg_H2_False_False_all_20_filterTrue_nFalse_exp200.pdf}
        \caption{Hydrogen price}
        \label{fig:hydrogen-price-hourly}
    \end{subfigure}
    \hfill
    
    \caption{Cost components of hydrogen electrolysis and the respective main influence factors on it {\bf with} hourly temporal matching. The CAPEX (\ref{fig:capex-rel-hourly}) depends on the electrolysis capacity factor (\ref{fig:ely-cf-hourly}), whereas the OPEX (\ref{fig:opex-rel-hourly}) is influenced by the price of electricity for electrolysis (\ref{fig:electricity-price-hourly}). With temporal hydrogen regulation, the cost of hydrogen  (\ref{fig:hydrogen-cost-hourly}) is not reflected by the price of hydrogen (\ref{fig:hydrogen-price-hourly}), which includes the additional constraint of hourly matching and hence the installation of RE and storage capacities.}
    \label{fig:costbreakdown_hourly}
\end{figure*}

\begin{figure*}[h!]
    \centering
    \begin{subfigure}[b]{0.45\linewidth}
        \centering
        \includegraphics[width=\linewidth]{../\runsensnogreenhy/graphics/integrated_comp/contour_capex_ely_rel_20_filterTrue_nFalse_exp200.pdf}
        \caption{CAPEX share of hydrogen costs}
        \label{fig:capex-rel-norule}
    \end{subfigure}
    \hfill
    \begin{subfigure}[b]{0.45\linewidth}
        \centering
        \includegraphics[width=\linewidth]{../\runsensnogreenhy/graphics/integrated_comp/contour_cf_electrolysis_20_filterTrue_nFalse_exp200.pdf}
        \caption{Electrolysis capacity factor}
        \label{fig:ely-cf-norule}
    \end{subfigure}
    \hfill
    \begin{subfigure}[b]{0.45\linewidth}
        \centering
        \includegraphics[width=\linewidth]{../\runsensnogreenhy/graphics/integrated_comp/contour_opex_ely_rel_20_filterTrue_nFalse_exp200.pdf}
        \caption{OPEX share of hydrogen costs}
        \label{fig:opex-rel-norule}
    \end{subfigure}
    \hfill
    \begin{subfigure}[b]{0.45\linewidth}
        \centering
        \includegraphics[width=\linewidth]{../\runsensnogreenhy/graphics/integrated_comp/contour_mg_AC_inclu_H2_El_all_20_filterTrue_nFalse_exp200.pdf}
        \caption{Price of electricity (electrolysis demand weighted)}
        \label{fig:electricity-price-norule}
    \end{subfigure}
    \hfill
    \begin{subfigure}[b]{0.45\linewidth}
        \centering
        \includegraphics[width=\linewidth]{../\runsensnogreenhy/graphics/integrated_comp/contour_lcoh_compo_20_filterTrue_nFalse_exp200.pdf}
        \caption{Total hydrogen cost}
        \label{fig:hydrogen-cost-norule}
    \end{subfigure}
    \hfill
    \begin{subfigure}[b]{0.45\linewidth}
        \centering
        \includegraphics[width=\linewidth]{../\runsensnogreenhy/graphics/integrated_comp/contour_mg_H2_False_False_all_20_filterTrue_nFalse_exp200.pdf}
        \caption{Hydrogen price}
        \label{fig:hydrogen-price-norule}
    \end{subfigure}
    \hfill
    
    \caption{Cost components of hydrogen electrolysis and the respective main influence factors on it {\bf without} temporal hydrogen regulation. The CAPEX (\ref{fig:capex-rel-norule}) depends on the electrolysis capacity factor (\ref{fig:ely-cf-norule}), whereas the OPEX (\ref{fig:opex-rel-norule}) is influenced by the price of electricity for electrolysis (\ref{fig:electricity-price-norule}). With temporal hydrogen regulation, the cost of hydrogen  (\ref{fig:hydrogen-cost-norule}) is not reflected by the price of hydrogen (\ref{fig:hydrogen-price-norule}), which includes the additional constraint of hourly matching and hence the installation of RE and storage capacities.}
    \label{fig:costbreakdown_norule}
\end{figure*}

\clearpage

\section{High export scenarios}
\label{sec:highexportsens}
\subsection{Electricity supply and demand}

\begin{figure*}[h!]
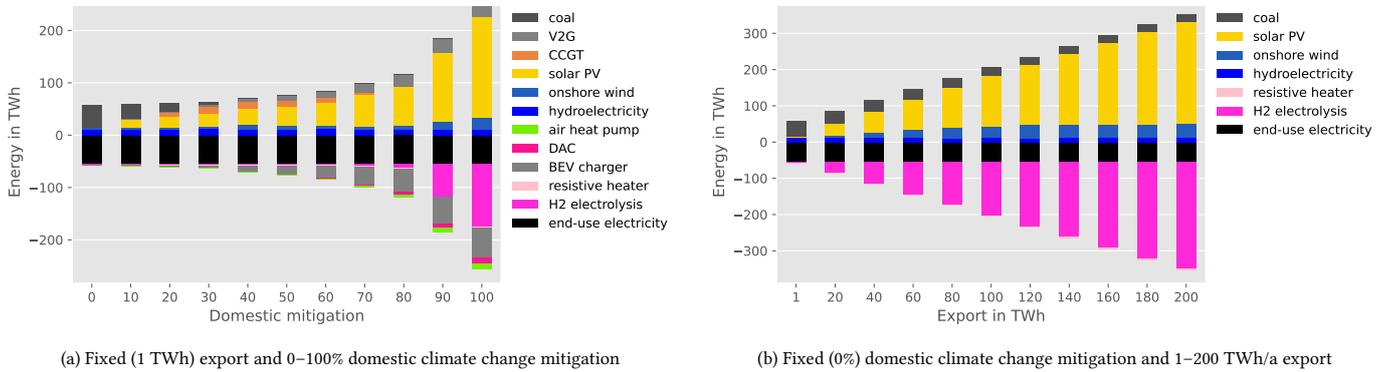

    \centering
    \begin{subfigure}[b]{0.49\linewidth}
        \centering
        \includegraphics[trim={0cm 0cm 0cm 1cm}, clip, width=\linewidth]{../workflow/subworkflows/pypsa-earth-sec/results/\runstandard/0exp-only/200/graphs/balances-AC.pdf}
        \caption{Fixed (1 TWh) export and 0--100\% domestic climate change mitigation}
        \label{fig:balances-ac-0exp}
    \end{subfigure}
    \hfill
    \begin{subfigure}[b]{0.49\linewidth}
        \centering
        \includegraphics[trim={0cm 0cm 0cm 1cm}, clip, width=\linewidth]{../workflow/subworkflows/pypsa-earth-sec/results/\runstandard/co2l20-only/200/graphs/balances-AC.pdf}
        \caption{Fixed (0\%) domestic climate change mitigation and 1--200 TWh/a export}
        \label{fig:balances-ac-co2l20-200}
    \end{subfigure}
    \hfill
    \caption{Electricity supply and demand at fixed export levels and increasing domestic climate change mitigation (\ref{fig:balances-ac-0exp}) and vice versa (\ref{fig:balances-ac-co2l20-200}). Increasing domestic climate change mitigation first phases out carbon-intensive coal generation in favor of CCGT, at medium to high domestic climate change mitigation the electricity system is fully renewable supported by flexibility through Vehicle-to-Grid (V2G) and sector coupling. Increasing electricity demands cover EVs and hydrogen generation for other sectors.
    At increasing hydrogen exports the additional electricity required for hydrogen electrolysis is covered by onshore wind and solar PV, as imposed by the temporal hydrogen regulation. 
    }
    \label{fig:balances-ac}
\end{figure*}

\subsection{Relative cost of domestic electricity and hydrogen export}

\begin{figure*}[h!]
    \centering
    \begin{subfigure}[b]{0.49\linewidth}
        \centering
        \includegraphics[width=\linewidth]{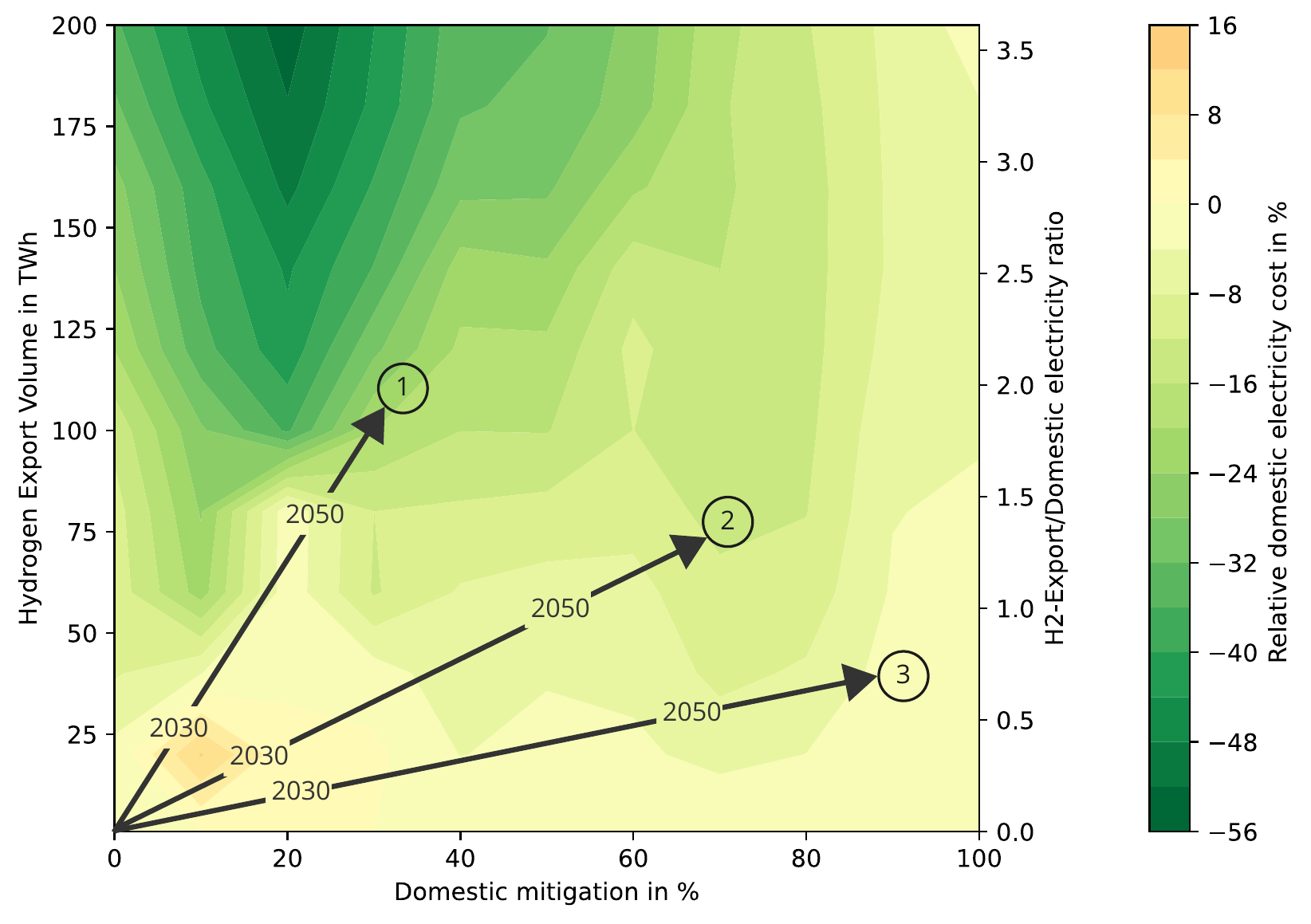}
        \caption{Rel. cost of electricity for domestic customers (normalized to 1 TWh/a hydrogen export)}
        \label{fig:expense_ac_200}
    \end{subfigure}
    \hfill
    \begin{subfigure}[b]{0.49\linewidth}
        \centering
        \includegraphics[width=\linewidth]{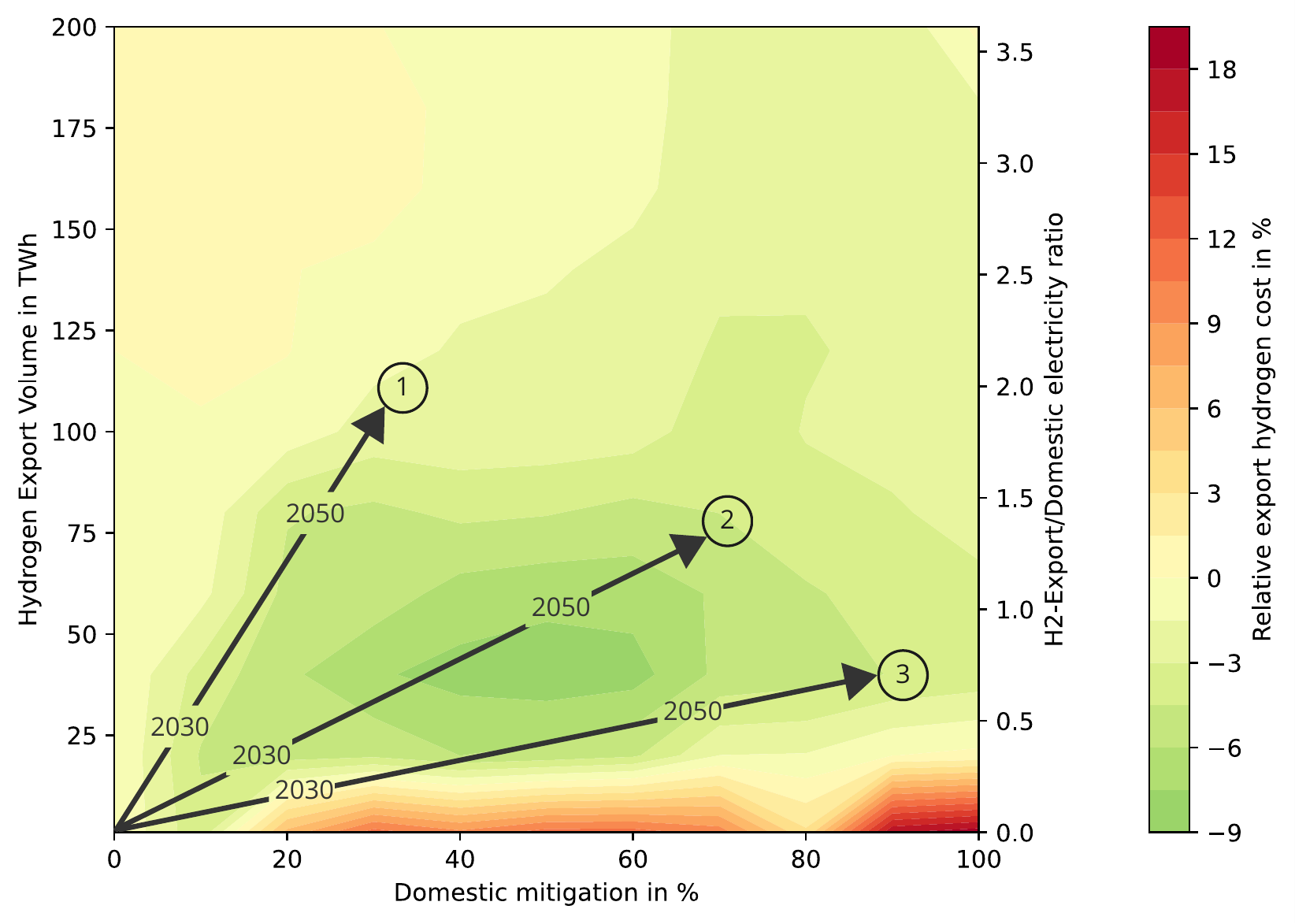}
        \caption{Rel. cost of hydrogen for exporters (normalized to 0\% \co reduction)}
        \label{fig:expense_h2_200}
    \end{subfigure}
    \hfill
    \caption{  
    Rel. cost for domestic electricity consumers (\ref{fig:expense_ac_200}) and hydrogen exporters (\ref{fig:expense_h2_200}),
    normalized to costs at 1 TWh/a hydrogen export (\ref{fig:expense_ac_200}) and
    to 0\% \co reduction (\ref{fig:expense_h2_200})
    at each domestic climate change mitigation level. Domestic electricity consumers profit from increasing hydrogen exports, especially at low domestic climate change mitigation and high exports. Hydrogen exporters profit from domestic climate change mitigation at medium mitigation efforts. Both (\ref{fig:expense_ac_200}) and (\ref{fig:expense_h2_200}) include possible pathways of i) quick exports and slow climate change mitigation, ii) balanced exports and mitigation and iii) slow exports and quick climate change mitigation.}
    \label{fig:expenses_default_200}
\end{figure*}

\subsection{Effect of temporal hydrogen regulation on electricity and hydrogen costs in all scenarios} 

\begin{figure*}[h!]
    \centering
    \begin{subfigure}[b]{0.49\linewidth}
        \centering
        \includegraphics[trim={0cm 0cm 0cm 0.65cm}, clip, width=\linewidth]{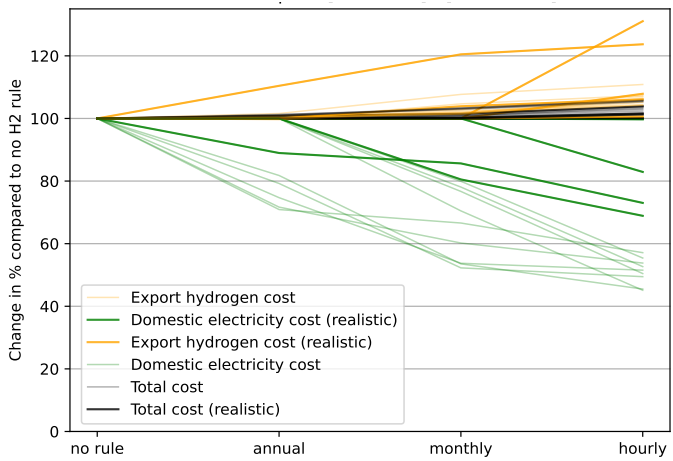}
        \caption{All scenarios up to 120 TWh}
        \label{fig:expenses_all_120}
    \end{subfigure}
    \hfill
    \begin{subfigure}[b]{0.49\linewidth}
        \centering
        \includegraphics[trim={0cm 0cm 0cm 0.65cm}, clip, width=\linewidth]{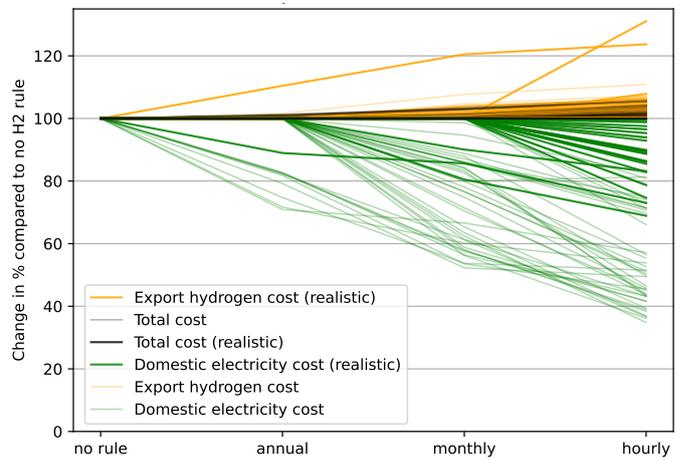}
        \caption{All scenarios up to 200 TWh}
        \label{fig:expenses_all_200}
    \end{subfigure}
    \hfill
    \caption{Supplement to Fig. \ref{fig:expenses_real_120}. Relative change of electricity and hydrogen cost and total system cost depending on the temporal hydrogen regulation. Domestic electricity consumers profit across all export and mitigation scenarios but most at high export and low climate change mitigation. Hydrogen exporters experience higher cost with stricter temporal hydrogen regulation. The temporal hydrogen regulation regulates the welfare distribution between both groups.}
    \label{fig:expenses_all}
\end{figure*}

\clearpage

\section{Carbon Management}
\label{subsec:carbon_mgmt}

\begin{figure*}[h!]
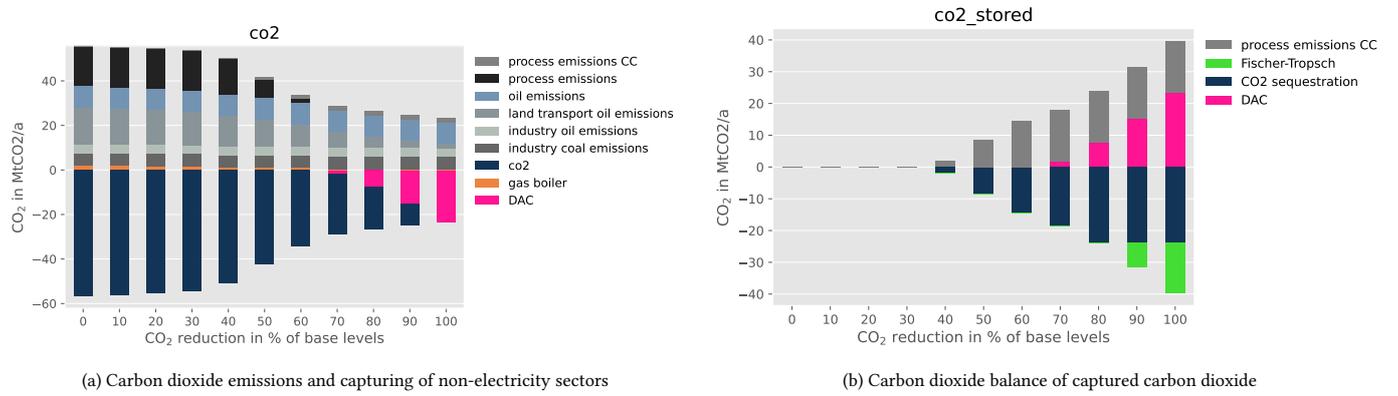

    \centering
    \begin{subfigure}[b]{0.49\linewidth}
        \centering
        \includegraphics[width=\linewidth]{../workflow/subworkflows/pypsa-earth-sec/results/\runstandard/0exp-only/graphs/balances-co2.pdf}
        \caption{Carbon dioxide emissions and capturing of non-electricity sectors}
        \label{fig:carbon-atmo}
    \end{subfigure}
    \hfill
    \begin{subfigure}[b]{0.49\linewidth}
        \centering
        \includegraphics[width=\linewidth]{../workflow/subworkflows/pypsa-earth-sec/results/\runstandard/0exp-only/graphs/balances-co2_stored.pdf}
        \caption{Carbon dioxide balance of captured carbon dioxide}
        \label{fig:carbon-store}
    \end{subfigure}
    \hfill
    \caption{Carbon dioxide \ref{fig:carbon-atmo} emissions and capturing as well as \ref{fig:carbon-store} management of captured carbon dioxide}
    \label{fig:carbon-mgmt}
\end{figure*}

Figure \ref{fig:carbon-mgmt} shows the carbon management of scenarios with increasing emission limit at 1 TWh/a export volume.

\clearpage
\section{System operation}

\subsection{Operation of electrolysers and Fischer-Tropsch in dependence of domestic climate change mitigation}

\begin{figure}[h]
    \centering
        \begin{subfigure}[h]{0.33\textwidth}
            \centering
        \includegraphics[width=\textwidth]{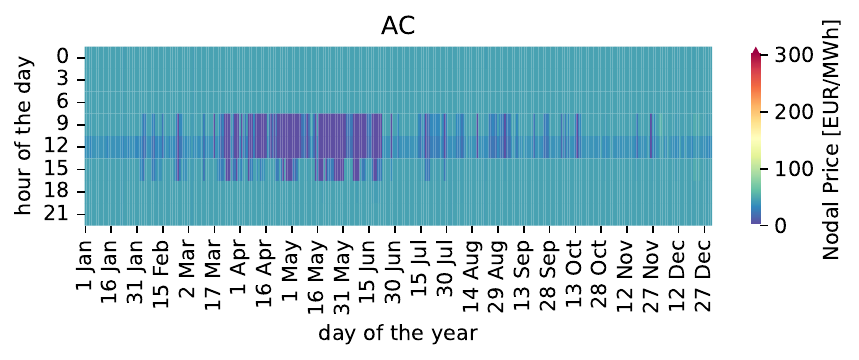}
    \end{subfigure}
    \begin{subfigure}[h]{0.33\textwidth}
        \centering
        \includegraphics[width=\textwidth]{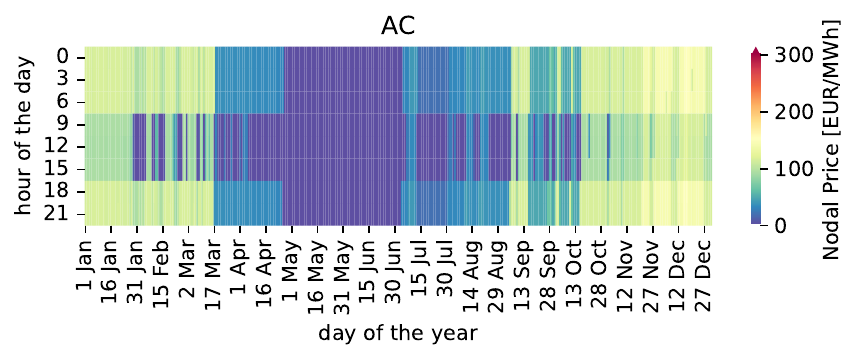}
    \end{subfigure}
    \begin{subfigure}[h]{0.33\textwidth}
        \centering
        \includegraphics[width=\textwidth]{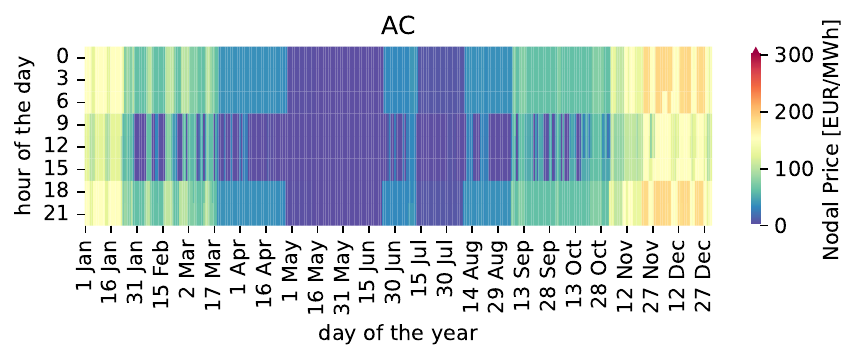}
    \end{subfigure}

    \begin{subfigure}[h]{0.33\textwidth}
        \centering
        \includegraphics[width=\textwidth]{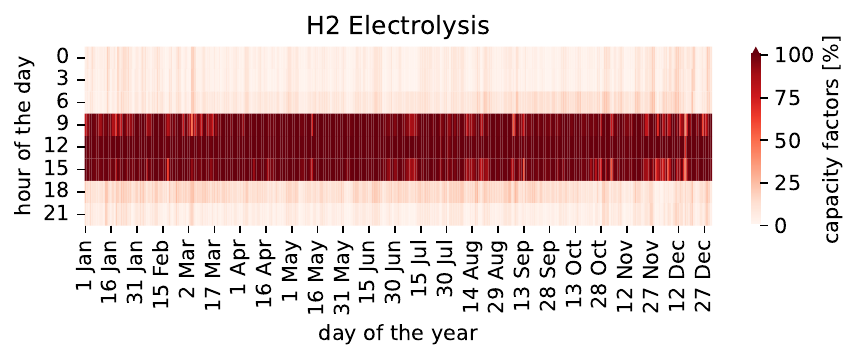}
    \end{subfigure}
    \begin{subfigure}[h]{0.33\textwidth}
        \centering
        \includegraphics[width=\textwidth]{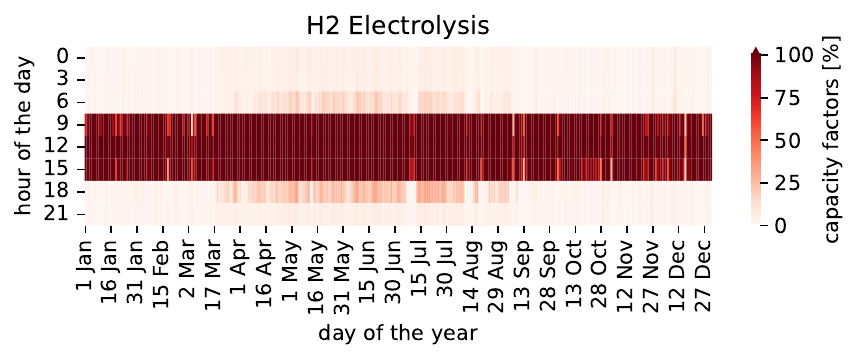}
    \end{subfigure}
    \begin{subfigure}[h]{0.33\textwidth}
        \centering
        \includegraphics[width=\textwidth]{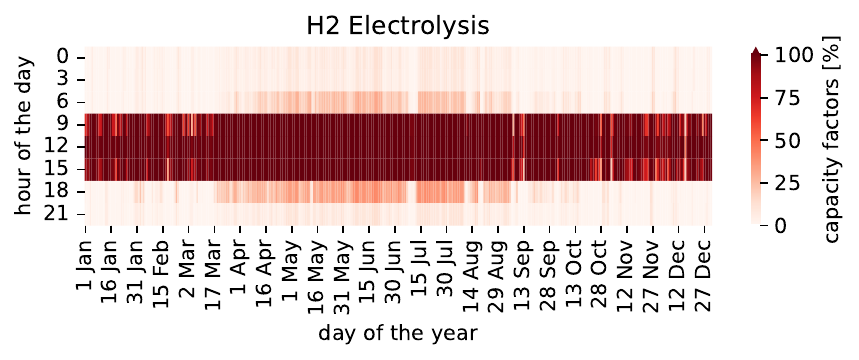}
    \end{subfigure}

    \begin{subfigure}[h]{0.33\textwidth}
        \centering
        \includegraphics[width=\textwidth]{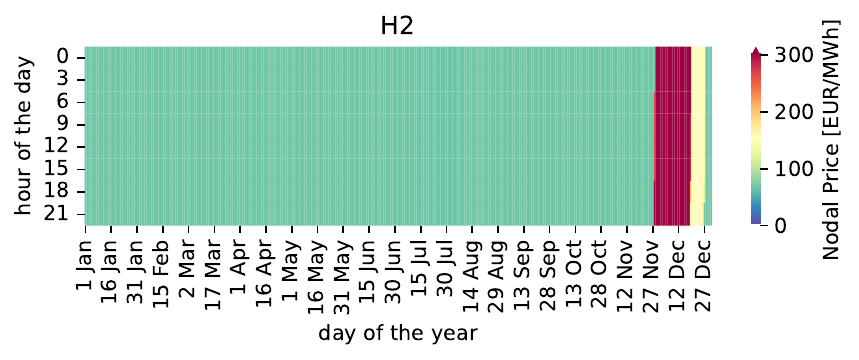}
    \end{subfigure}
    \begin{subfigure}[h]{0.33\textwidth}
        \centering
        \includegraphics[width=\textwidth]{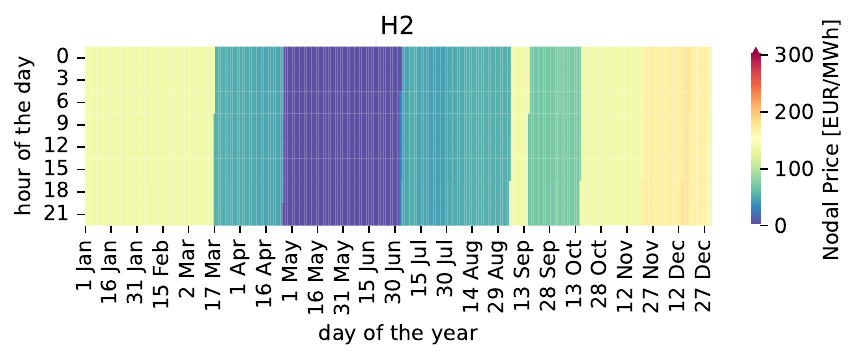}
    \end{subfigure}
    \begin{subfigure}[h]{0.33\textwidth}
        \centering
        \includegraphics[width=\textwidth]{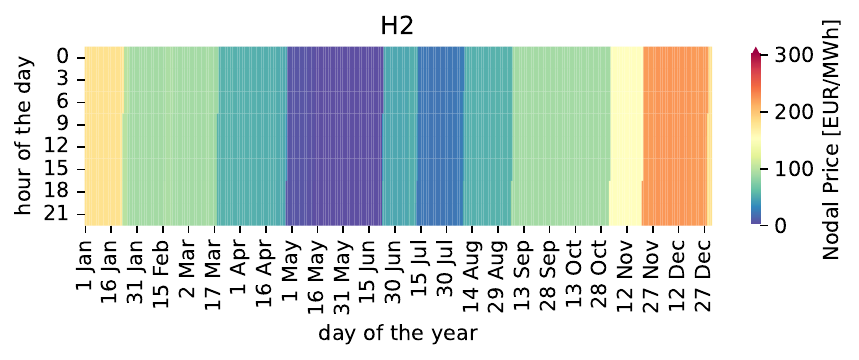}
        
    \end{subfigure}

    \begin{subfigure}[h]{0.33\textwidth}
        \centering
        \includegraphics[width=\textwidth]{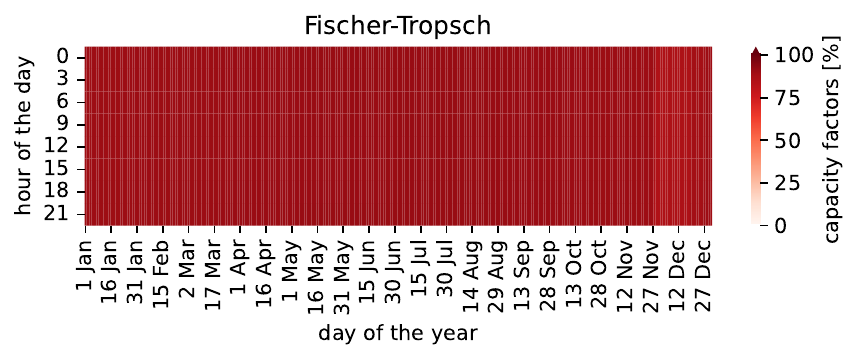}
        \caption{20\% climate change mitigation}
        \label{fig:operation20}
    \end{subfigure}
    \begin{subfigure}[h]{0.33\textwidth}
        \centering
        \includegraphics[width=\textwidth]{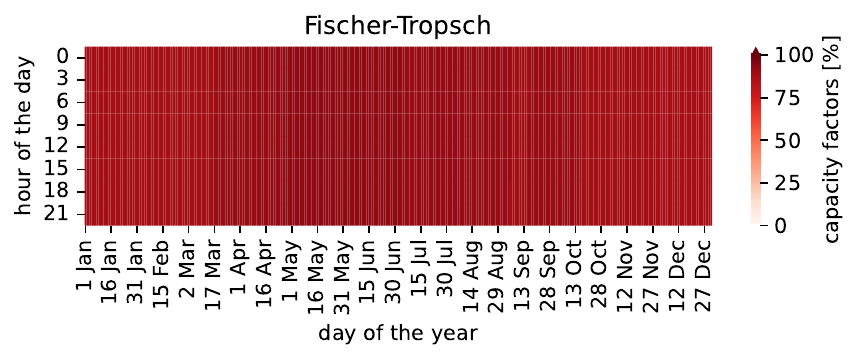}
        \caption{70\% climate change mitigation}
        \label{fig:operation70}
    \end{subfigure}
    \begin{subfigure}[h]{0.33\textwidth}
        \centering
        \includegraphics[width=\textwidth]{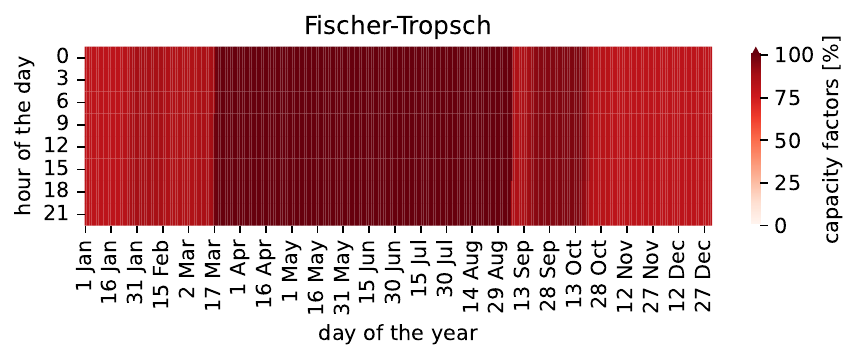}
        \caption{90\% climate change mitigation}
        \label{fig:operation90}
    \end{subfigure}

    \caption{Operation of electrolysers/Fischer-Tropsch and prices of electricity and hydrogen at 40 TWh/a export and 20\%, 70\% and 90\% domestic climate change mitigation. The electrolyser operation depends on the electricity prices, strong diurnal electricity price patterns as in \ref{fig:operation70} result in a corresponding electrolyser operation. Increased battery capacity at \ref{fig:operation90} results in smoother electricity prices also increase the annual capacity factor of the electrolyser. Note: electricity and hydrogen prices are spatially averaged in non-weighted manner, graph style adapted from Neumann et al. \cite{Neumann2022}}
    \label{fig:operation-ely-ft}
\end{figure}

\subsection{Capacity factors of electrolysis and Fischer-Tropsch}

\begin{figure*}[h] %
    \centering
    \begin{subfigure}[b]{0.49\linewidth}
        \centering
        \includegraphics[width=\linewidth]{graphics/integrated_comp/contour_cf_electrolysis_20_filterTrue_nFalse_exp200.pdf}
        \caption{Capacity factor of electrolysis}
        \label{fig:cf-ely}
    \end{subfigure}
    \hfill
    \begin{subfigure}[b]{0.49\linewidth}
        \centering
        \includegraphics[width=\linewidth]{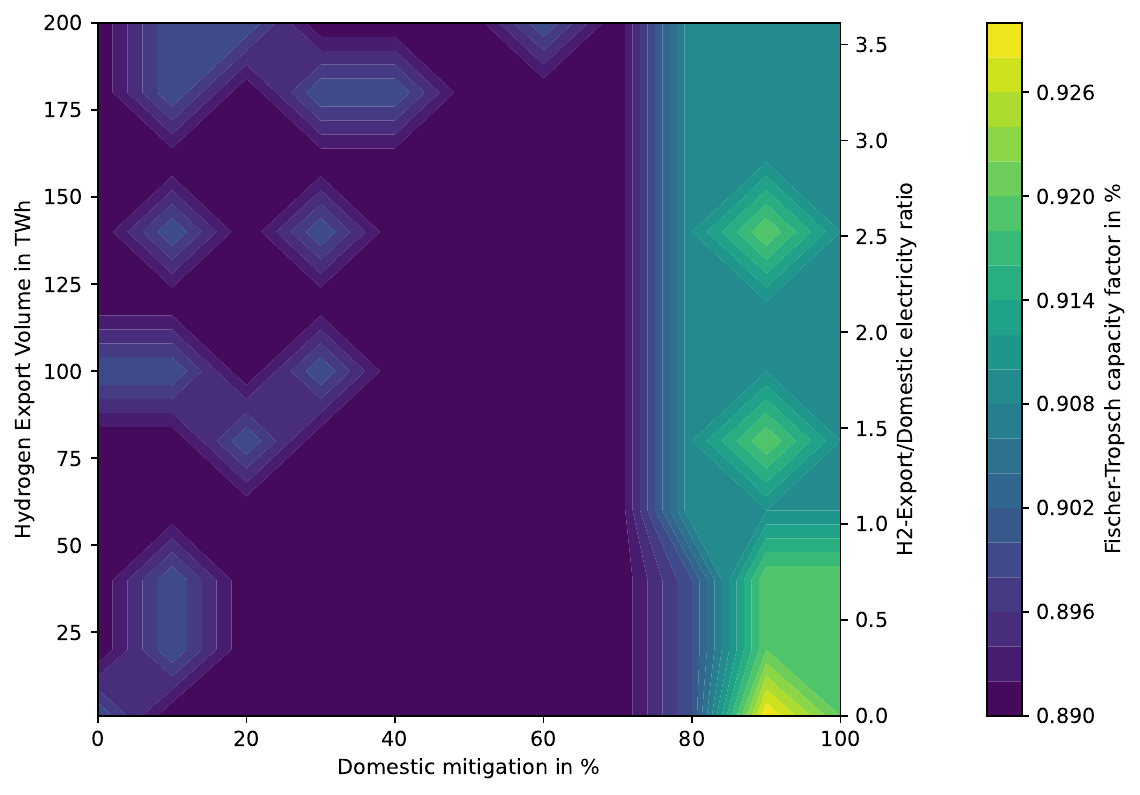}
        \caption{Capacity factor of Fischer-Tropsch}
        \label{fig:cf-ft}
    \end{subfigure}
    \hfill
    \caption{Capacity factors of electrolysis and Fischer-Tropsch. The electrolysis has a high capacity factor in low export and low domestic climate change scenarios. The capacity factor for Fischer-Tropsch is close to 1 given the limited flexibility of the process.}
    \label{fig:cf-ely-ft}
\end{figure*}

\clearpage

\addcontentsline{toc}{section}{Supplementary References}
\renewcommand{\ttdefault}{\sfdefault}

\end{document}